\documentclass[apj]{emulateapj}
\usepackage{natbib}

\usepackage{amsmath}
\usepackage{apjfonts}
\usepackage{multirow}
\usepackage{graphicx}
\usepackage{amssymb}
\usepackage{float}
\usepackage{longtable}

\usepackage{CJK}

\shorttitle{Merging Galaxies with Tidal Tails to $z=1$}
\shortauthors{Wen \& Zheng}

\begin{document}
\begin{CJK*}{UTF8}{gbsn}

\title{Merging Galaxies with Tidal Tails  in COSMOS to $z=1$}
\author{Zhang~Zheng~Wen (闻璋正) \altaffilmark{1,2} and Xian~Zhong~Zheng (郑宪忠)\altaffilmark{1}}
\altaffiltext{1}{Purple Mountain Observatory, Chinese Academy of Sciences, 2 West Beijing Road, Nanjing, 210008, China; zzwen@pmo.ac.cn, xzzheng@pmo.ac.cn}
\altaffiltext{2}{University of Chinese Academy of Sciences, 19A Yuquan Road Beijing, China}

\begin{abstract}
Tidal tails are created in major mergers involving disk galaxies. How the tidal tails trace the assembly history of massive galaxies remains to be explored.
We identify a sample of 461 merging galaxies with long tidal tails from 35\,076 galaxies mass-complete at $M_\star\ge 10^{9.5}\,M_{\odot}$ and $0.2\leq z\leq1$ based on \textit{HST}/ACS F814W imaging data and public catalogs of the COSMOS field. The long tails refer to these with length equal to or longer than the diameter of their host galaxies. The mergers with tidal tails are selected using our novel $A_{\rm O}-D_{\rm O}$ technique for strong asymmetric features together with visual examination. Our results show that the fraction of tidal-tailed mergers evolves mildly with redshift, as $\sim (1+z)^{2.0\pm0.4}$, and becomes relatively higher in less massive galaxies out to $z=1$. With a timescale of 0.5\,Gyr for the tidal-tailed mergers, we obtain that the occurrence rate of such mergers follows $0.01\pm 0.007\,(1+z)^{2.3\pm 1.4}$\,Gyr$^{-1}$ and corresponds to $\sim0.3$ events since $z=1$ and roughly one-third of the total budget of major mergers from the literature. For disk-involved major mergers, nearly half of them have undergone a phase with long tidal tails. 
 \end{abstract}


\section{INTRODUCTION} \label{sec:sec1}

The stellar mass growth of galaxies is driven by in-situ star formation and mergers.  Quantifying the relative roles of the two processes across cosmic time is one central task in extragalactic astronomy, and provides key observational constraints on theoretical models of galaxy formation and evolution. The in-situ star formation on galactic scales has been much explored out to high redshift in either the volume-averaged or the sub-population statistic manner. The cosmic star formation rate (SFR) density peaks at $z\sim2-3$ and declines rapidly since then \citep{1996ApJ...460L...1L,1996MNRAS.283.1388M,2011ARA&A..49..525S,2014ARA&A..52..415M}. The majority of star formation takes place in the main-sequence galaxies which form a tight correlation between SFR and stellar mass \citep[e.g.,][]{2015A&A...575A..74S}, and mostly hosted by those in the mass regime around $M^\star$ \citep{2007ApJ...661L..41Z,2011ApJ...730...61K}. The main sequence relation exists among star-forming galaxies (SFGs) out to high $z$ \citep[e.g.,][]{2014ApJS..214...15S}, and its slope and dispersion are shaped by physical processes related to bulge growth \citep{2015ApJ...808L..49G,2015ApJ...811L..12W}. The decline of the global SFR density is coupled with the rapid increase of stellar mass density of quiescent galaxies since $z\sim 3$ \citep{ 2013A&A...556A..55I,2014ApJ...783...85T} continuously fed by SFGs through quenching star formation \citep[e.g.,][]{2007ApJ...663..834B,2007ApJ...665..265F,2013ApJ...767...50M,2014MNRAS.438.1870D,2015MNRAS.451.2933B}.
In contrast, galaxy merging is not only an important process regulating the assembly of galaxies, but also a key mechanism in triggering starburst, shaping galaxy structure, igniting active galactic nucleus (AGN), and ceasing further star formation through exhausting gas or AGN/supernova feedback \citep[e.g.,][]{1996ApJ...464..641M,2005Natur.433..604D,2005MNRAS.361..776S,2005ApJ...620L..79S,2006MNRAS.365...11C,2006ApJS..163....1H,2008ApJS..175..390H}. 




Galaxy merger rate as a function of redshift is one of fundamental relationships to delineate the hierarchical growth history of galaxies and cosmic structures \citep[e.g.,][]{1978MNRAS.183..341W,2008MNRAS.391..481S}. On the theoretical side, $N$-body simulations predict that merger rate of dark matter halos strongly evolves with redshift as $(1+z)^m$ and $m=2\sim3$ \citep{2001ApJ...546..223G,2008MNRAS.386..577F,2009ApJ...701.2002G,2010MNRAS.406.2267F}, with an uncertainty about a factor of 2 \citep{2010ApJ...724..915H}. Conversion of the halo merger rate into galaxy merger rate has to rely on assumptions of baryonic physics adopted in the theoretical models of galaxy evolution such as the sub-halo structures and mass functions, halo occupation statistics, gas hydrodynamics and cooling, star formation and feedback (\citealt{2009ApJ...697.1971J,2010ApJ...724..915H} for reference therein). On the observational side, measurements of galaxy merger rate from morphological studies and pair studies are not consistent with each other \citep{2011ApJ...742..103L}. The discrepancy arises from uncertainties in sample selection, identification of mergers, correction for false mergers, incompleteness with redshift, cosmic variance and uncertainty in estimation of merger timescale.


Galaxy merger rate is estimated through counting either close pairs \citep[e.g.,][]{2007ApJS..172..320K,2008ApJ...681..232L,2009MNRAS.394L..51B,2009A&A...498..379D,2009ApJ...697.1369B,2011ApJ...738L..25W,2012ApJ...747...34B,2012ApJ...744...85M,2012ApJ...746..162N,2012ApJ...747...85X,2014ApJ...795..157K,2015A&A...576A..53L} or morphology-disturbed galaxies.
For the latter, visual examination and non-parametric methods are widely used to select galaxy major mergers, which often exhibit tidal features like shells, arcs, plumes, rings, streams, tails and double cores. Such disturbed features can be caught in visual inspection \citep[e.g.,][hereafter B10]{2009ApJ...697.1971J,2010ApJ...709.1067B}.
The non-parametric measurements are to quantify morphological parameters, including  the rotational asymmetry $A$ \citep{2003ApJS..147....1C}, Gini-$M_{20}$ \citep{2004AJ....128..163L} or $A_{\rm O}-D_{\rm O}$ \citep{2014ApJ...787..130W}, and separate mergers from those with regular morphologies in the corresponding parametric space. However, morphology studies largely depend on the quality of imaging data as cosmic dimming effect may cause redshift-dependent biases. Noise contamination strongly affects the selection of mergers \citep{2003ApJS..147....1C,2008MNRAS.386..909C,2009MNRAS.394.1956C,2009A&A...501..505L,2009ApJ...697.1764S}. Similarly, the Gini coefficient is also influenced by the adopted aperture for a galaxy \citep{2008ApJS..179..319L}.
Another uncertainty in estimating merger rate is related to detection timescale. The averaged  timescale for detection of asymmetric features ($A$) is strongly correlated with the gas fraction and mass ratio of the merger progenitors \citep{2008MNRAS.391.1137L,2010MNRAS.404..575L,2010MNRAS.404..590L,2011ApJ...742..103L}. The contamination of false mergers with peculiar morphologies \citep{2009ApJ...706.1364F,2011ApJ...731...65F} may also account for a significant fraction of the merger candidates preliminarily selected by the non-parameter methods \citep[e.g.,][]{2012MNRAS.419.2703H}.

One way to improve the measurements of merger rate is to narrow the selection of galaxy mergers down to a specific type or merging stage so that the related uncertainties can be accordingly reduced \citep[e.g.,][]{2014AJ....148..137L}. Tidal tails are created in major mergers involving at least one disk galaxy and can be used as a secure sign of major merger event \citep{1972ApJ...178..623T}. Moreover, the tidal tails contain stars and gas, allowing dwarf galaxies and star clusters to form, and contributing to the intergalactic medium \citep[see][and references therein]{2013LNP...861..327D}. The length and orientation of tidal tails turn out to be regulated by the inclination of progenitor disk galaxies to the approaching orbit. Tidal tails thus trace the assembly history of galaxies with disk progenitors and reflect the alignment of progenitor galaxies in the cosmic web \citep{2006MNRAS.369.1293Y,2009ApJ...706..747Z,2013ApJ...779..160Z}.

Yet, few works made use of tidal tails as the probe for major mergers to measure merger rate (B10). Tidal tails are faint structures and usually contribute only limited fraction of the total light of a merging system. The morphological parameters rotational asymmetry $A$,  Gini and $M_{20}$ are likely insensitive to probe of such faint structures in a galaxy because these parameters are flux weighted and usually dominated by luminous structures of the galaxy.
\citet{2014ApJ...787..130W} developed a novel $A_{\rm O}-D_{\rm O}$ method to select disturbed morphologies through quantifying the outer asymmetry ($A_{\rm O}$) and the deviation of the intensity weighted centroids of the outer half light region and the inner half light region ($D_{\rm O}$) of a galaxy. This technique is efficient in detecting tidal-tailed galaxies (see \citealt{2014ApJ...787..130W} for more details). Galaxies of different morphological types fall on a tight sequence in the $A_{\rm O}$ versus $D_{\rm O}$ diagram and the tailed galaxies usually have larger $A_{\rm O}$ and $D_{\rm O}$ relative to those with regular morphologies.


We carry out a search for merging galaxies with long tidal tails in the COSMOS field. The \textit{HST}/ACS F814W ($I$) imaging observations of the COSMOS survey \citep{2007ApJS..172....1S,2007ApJS..172..196K} cover $\sim1.5$ deg$^2$ sky area, providing high-resolution images to probe faint structures such as tidal tails (in a size of tens kpc) for a large sample of distant galaxies at $z\le1$. This paper is arranged as following:  a brief description of the data and sample selection is presented in Section~2; the $A_{\rm O}-D_{\rm O}$ method is given in Section~3. Meanwhile, we perform simulations to test $A_{\rm O}$ and $D_{\rm O}$ with surface brightness dimming included. In Section~4, we present the selected merging galaxies with tidal tails using the $A_{\rm O}-D_{\rm O}$ method. Finally, we report our merger fraction and merger rate over $0.2\leq z \leq1$ in Section~5. Throughout this paper, we adopt a concordance cosmology with $H_{0}=70$\, km\,s$^{-1}$\,Mpc$^{-1}$, $\Omega_m=0.3$ and $\Omega_\Lambda=0.7$. All photometric magnitudes are given in the AB system.

\section{Data and Sample Selection} \label{sec:sec2}

\subsection{The Data}

Our study utilizes the public data and catalogs from multi-band deep surveys of the COSMOS field.
The UltraVISTA survey \citep{2012A&A...544A.156M} provides ultra-deep near-IR imaging observations of this field in the $Y,\, J,\, H$ and $K_s$-band as well as a narrow band (NB118) \citep{2013A&A...560A..94M}. The UltraVISTA photometric catalogs (v4.1) are available to the public, giving in total 154\,803 sources based on $K_s$-band detection down to 5\,$\sigma$=23.4\,mag with 90\% completeness for point sources (see \citealt{2013ApJS..206....8M} for more details). The PSF-matched photometry with 30 photometric bands from 0.15 to 24\,$\micron$, stellar mass and redshift catalogs are also provided. Stellar masses are determined using the FAST SED fitting code \citep{2009ApJ...700..221K} while photometric redshifts are derived with the EAZY code \citep{2008ApJ...686.1503B}. Spectroscopic redshifts from the zCOSMOS 10k bright sample \citep{2009ApJS..184..218L} are also adopted. The photometric redshifts reach an accuracy of $\delta z/(1+z)=0.013$ for $z\lesssim 1.5$. Spectroscopic redshifts are first used if available.

The \textit{HST}/ACS $I$-band imaging data are publicly available, allowing us to measure  morphologies in the rest-frame optical for galaxies at $z\le1$. The \textit{HST}/ACS $I$-band images reach a 5\,$\sigma$ depth of 27.2 magnitude for point sources. The reduced images have a pixel scale of $0\farcs03$ and a point spread function (PSF) of  Full Width at Half Maximum $0\farcs09$, corresponding to a physical size of 332\,pc at $z=0.2$ and 805\,pc at $z=1$. Following the source detection configurations given in \citet{2007ApJS..172..219L} and \citet{2008ApJS..174..136C}, we run the software tool SExtractor \citep{1996A&AS..117..393B} to obtain segmentation maps for detected sources. The photometric catalogs of the \textit{HST}/ACS $I$-band images are also used for source deblending. The segmentation maps will be used to perform morphological analysis. We set the detection threshold to be 0.8\,$\sigma$ of background noise level in order to detect extended faint tidal structures (see \citealt{2014ApJ...787..130W} for more details).

\subsection{Sample Selection}


We intend to identify merging galaxies with long tidal tails from a sufficiently large sample of galaxies out to high $z$.
The parent galaxy sample needs to be complete in stellar mass because galaxy mergers often trigger starbursts and increase extinction via condensing interstellar medium \citep{1996A&A...312..397G,1996A&A...313..377D,2001ApJ...550..212B,2003ApJS..149..289B,2008AJ....136.1866K}.
The UltraVISTA catalog based on $K_s$-band selection is less affected by dust extinction and starburst, and thus shows a higher completeness for low-mass galaxies compared to the optical selection. It has been shown that the  UltraVISTA catalog is highly complete (completeness $>$95\%) for galaxies with stellar mass  $\log (M_\star/M_\odot)\ge 9.5$  at $z\leq 1$ \citep{2013ApJS..206....8M}.  We limit sample galaxies at 0.2$\leq z \leq$1 where \textit{HST} $I$-band images allow to examine galaxy morphologies in the rest-frame optical.  Finally, we select a sample of 35\,076 galaxies  with $\log (M_\star/M_\odot)\ge 9.5$ and $0.2\le z\le1$ from the UltraVISTA catalog of COSMOS.   We measure morphological parameters $A_{\rm O}$ and $D_{\rm O}$ from \textit{HST} images and visually examine the selected merging galaxies for apparent tidal tails.
The parent sample is divided  into four mass bins and four redshift bins to address the relationships of the merger rate with stellar mass and redshift (e.g., \citealt{2004ApJ...617L...9L,2005ApJ...625..621B,2009A&A...498..379D};B10).



\section{The  $A_{\rm O}-D_{\rm O}$ Selection for Merging Galaxies}

We utilize the $A_{\rm O}-D_{\rm O}$ method by \citet{2014ApJ...787..130W} to select galaxies with asymmetric features in galaxy outskirts, including tidal tails.  Here we revisit the parameter $A_{\rm O}$ with an improved noise correction.  Simulations are performed to examine how the cosmic dimming effect influences the selection of merging galaxies using the $A_{\rm O}$ and $D_{\rm O}$ parameters.

\subsection{Morphological Parameters $A_{\rm O}$ and $D_{\rm O}$}\label{sec:Ao}

The two structural parameters $A_{\rm O}$ and $D_{\rm O}$ presented in \citet{2014ApJ...787..130W} give quantitative measures of galaxy structures in the outskirts.  The outskirts of a galaxy refers to the outer half-light region (OHR) divided by an elliptical aperture from the inner half-light region (IHR) of the galaxy.
The outer asymmetry $A_{\rm O}$ measures the asymmetry of the OHR; and the outer centroid deviation $D_{\rm O}$, measures the deviation (or offset) between the flux-weighted centroids of the IHR and the OHR.

In practice, the half-light aperture is derived from the isophotal analysis of a galaxy image.  The surface brightness profile  is counted down to the threshold of 0.8\,$\sigma$ background rms  in order to optimize the detection of faint structures in the outskirts  \citep{2014ApJ...787..130W} .
We realize that the noise in the low-surface brightness regions bias the estimate of $A_{\rm O}$. We thus correct for the bias and revise the definition of   $A_{\rm O}$ as below:
         \begin{equation} \label{equ:eqAo}
         A_{\rm O}=\frac{\sum{|I_{0}-I_{180}|}-\delta_{2}}{ \sum{|I_{0}|}-\delta_{1}},
         \end{equation}
where   $\delta_{1}=f_1*\sum{|B_{0}|}$,  $f_1=N_{flux<1\,\sigma}/N_{\rm all}$,
$\delta_{2}=f_2*\sum{|B_{0}-B_{180}|}$, and $f_2=N_{|flux|<\sqrt{2}\,\sigma}/N_{\rm all}'$.
Here $I_{0}$ refers to the light distribution of the OHR of a galaxy image and $I_{180}$ represents the $180\arcdeg$-rotated version of $I_{0}$. Likewise, $B_{0}$ is a patch in the background of the image with the same shape as $I_{0}$.  Similarly, $B_{180}$ is the $180\arcdeg$-rotated $B_{0}$.
$\delta_{1}$ and $\delta_{2}$ are corrections for the noise contributions to the flux image $I_{0}$ and the residual image $I_{\rm 0}-I_{180}$, respectively. $f_{1}$ is the number fraction of pixels in the OHR dominated by noise; and $f_{2}$ is the number fraction of the OHR pixels dominated by noise in the residual image. $N_{\rm all}$ represents the total number of pixels in the OHR and $N_{\rm all}'$ is the total number of pixels of the residual. Parameter $\sigma$ is the standard deviation of noise in $I_{0}$. In \citet{2014ApJ...787..130W}, the centroid of the OHR is used as the rotational center. But in this work the centroid of the whole galaxy is adopted as the rotational center. More technical details and testing results can be found in the Appendix.

The outer centroid deviation $D_{\rm O}$ is calculated using the equation
\begin{equation} \label{equ:eqDo}
D_{\rm O}=\frac{\sqrt{(x_{\rm O}-x_{\rm I})^2+(y_{\rm O}-y_{\rm I})^2}}{R_{\rm e}},
\end{equation}
  where ($x_{\rm I}$, $y_{\rm I}$) and ($x_{\rm O}$, $y_{\rm O}$) refer to the positions of flux-weighted centroids of the IHR and the OHR in a galaxy image, respectively. The deviation is normalized to the effective radius of the galaxy $R_{\rm e}$, which is calculated using $R_{\rm e}=\sqrt{n_{\rm I}/\pi}$ and $n_{\rm I}$ is the area of the IHR in units of pixel count. No noise correction is introduced to $D_{\rm O}$ because background noise does not significantly modify the position of centroid of a galaxy as well as the effective radius.

      Galaxies of different morphologies lie on a tight sequence in the $A_{\rm O}-D_{\rm O}$ diagram  and those exhibiting a higher degree of disturbance in morphology have statistically higher $A_{\rm O}$ and  $D_{\rm O}$.  The merging galaxies with long tidal tails show high $A_{\rm O}$ and  $D_{\rm O}$.
      The $A_{\rm O}-D_{\rm O}$method has been proved to be effective in selecting tailed galaxies, compared with CAS or Gini-$M_{\rm 20}$ selection techniques \citep{2014ApJ...787..130W}.

   \subsection{Application to High-$z$ Galaxies}\label{sec:simulation}

\begin{figure}
\includegraphics[width=0.45\textwidth]{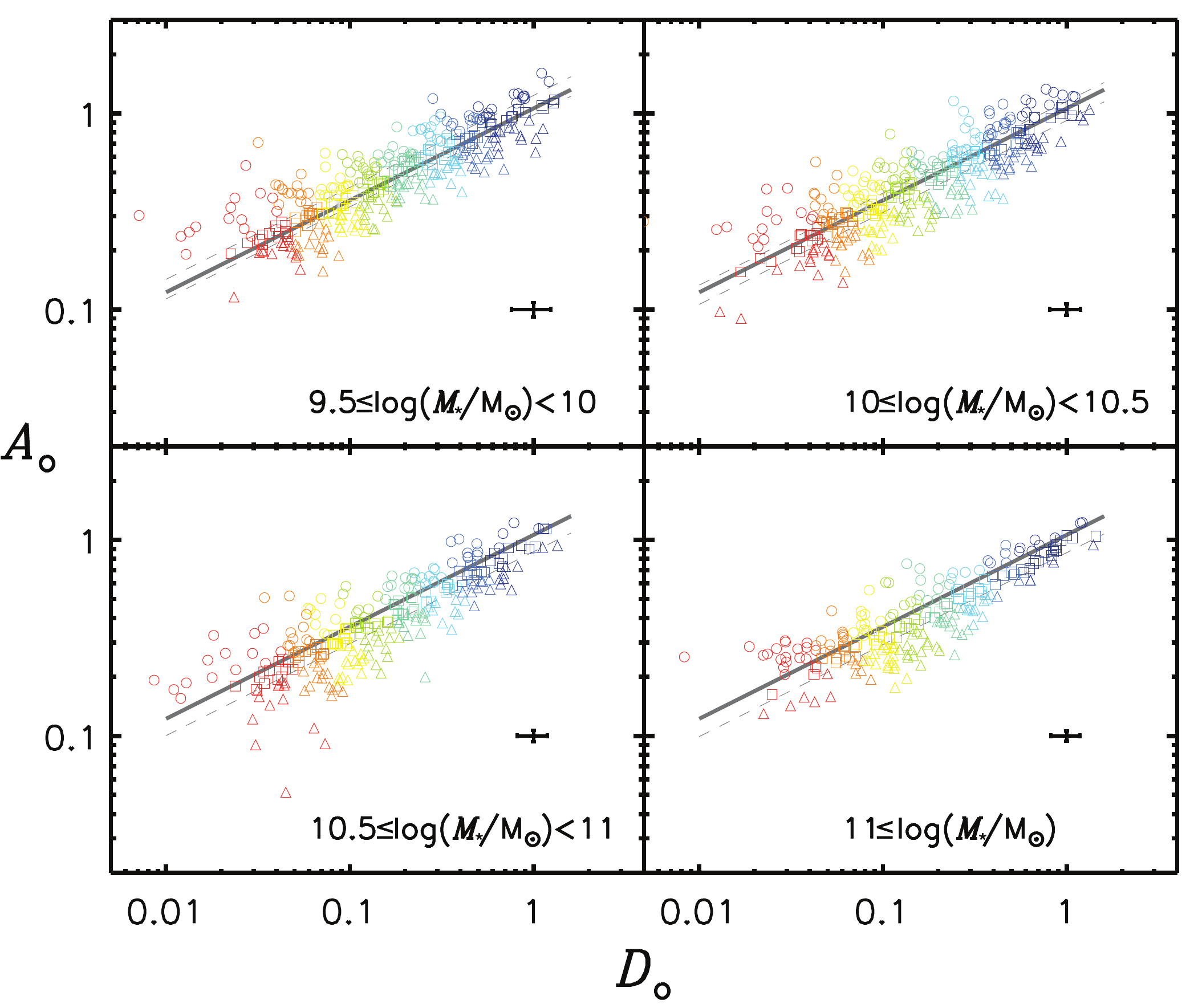}
\caption{Diagram of $A_{\rm O}$ versus $D_{\rm O}$  for 1178 galaxies at $0.2\le z<0.4$ from COSMOS. These galaxies form a sequence best-fitted by the solid line. Color coding denotes the 8 bins along the sequence and symbols marks 3 bins vertical to the sequence separated by the dashed lines. The typical 1\,$\sigma$ uncertainty in the measurements is marked in each panel.}
\label{fig:figA}
\end{figure}

     We aim to select merging galaxies out to $z=1$ using the $A_{\rm O}-D_{\rm O}$ technique. High-$z$ galaxies appear to be faint and less resolved compared to low-$z$ galaxies. We try to quantitatively estimate these effects on detection of asymmetric structures in galaxy outskirts by measuring the two structural parameters $A_{\rm O}$ and $D_{\rm O}$ for a sample of low-$z$ galaxies artificially redshifted to high $z$. We make use of the 4419 galaxies with $M_{\star}\ge10^{9.5} M_{\odot}$ in the lowest redshift bin $0.2\le z<0.4$ to carry out our test.
As shown in Figure~\ref{fig:figA}, these galaxies are sub-grouped into 24 bins in the $A_{\rm O}-D_{\rm O}$ space:  8 evenly-split bins along the sequence coded by different colors and 3 bins vertical to the sequence marked by different symbols. Objects are divided equally into the three bins vertical to the sequence. We then randomly select 15 galaxies in each bin, yielding 1178 at $0.2\le z<0.4$ with $M_{\star}\ge10^{9.5} M_{\odot}$ for further simulations. This sample is representatively spread in the $A_{\rm O}-D_{\rm O}$ space and thus able to trace the systematic effects when the sample galaxies are artificially redshifted to high $z$ with resolution and dimming effects counted. This galaxy sample is only used for testing measurements of structural parameters and thus unnecessary to be strictly complete in terms of selections for high-$z$ galaxies.

    We use \textit{HST}/ACS $I$-band science images of the selected 1178 galaxies  with $M_{\star}\ge10^{9.5} M_{\odot}$ at $0.2\le z<0.4$ to construct galaxy model images.  These model images are then used to simulate the images of high-$z$ galaxies and test the recovery of $A_{\rm O}$ and $D_{\rm O}$.
    Firstly, we apply median filtering to the \textit{HST} images of the selected galaxies to suppress the noise in the images.  A box of $6\times6$ pixels is adopted in the filtering, corresponding to 2 times FWHM of the \textit{HST}/ACS $I$-band PSF ($\sim 0\farcs09$). A group of continuous adjacent pixels 0.8\,$\sigma$ brighter than the background level in the original science image of a galaxy are taken as  the galaxy's pixels. Here $\sigma$ is the rms of the background noise.  These pixels in the smoothed image are subtracted by 0.8\,$\sigma$ to smooth the light profile at the edges and the rest pixels of the smoothed image are set to zero.
    Secondly,  galaxy model images are scaled to simulate galaxies redshifted to $z\sim0.3, 0.5$ and 0.9, respectively. This is done by decreasing the size and surface brightness of the model galaxies and adding background noise into the images.   Finally,  $A_{\rm O}$ and $D_{\rm O}$ are derived from the images of the simulated high-$z$ galaxies.

\begin{figure}[]
\centering
\includegraphics[width=0.45\textwidth]{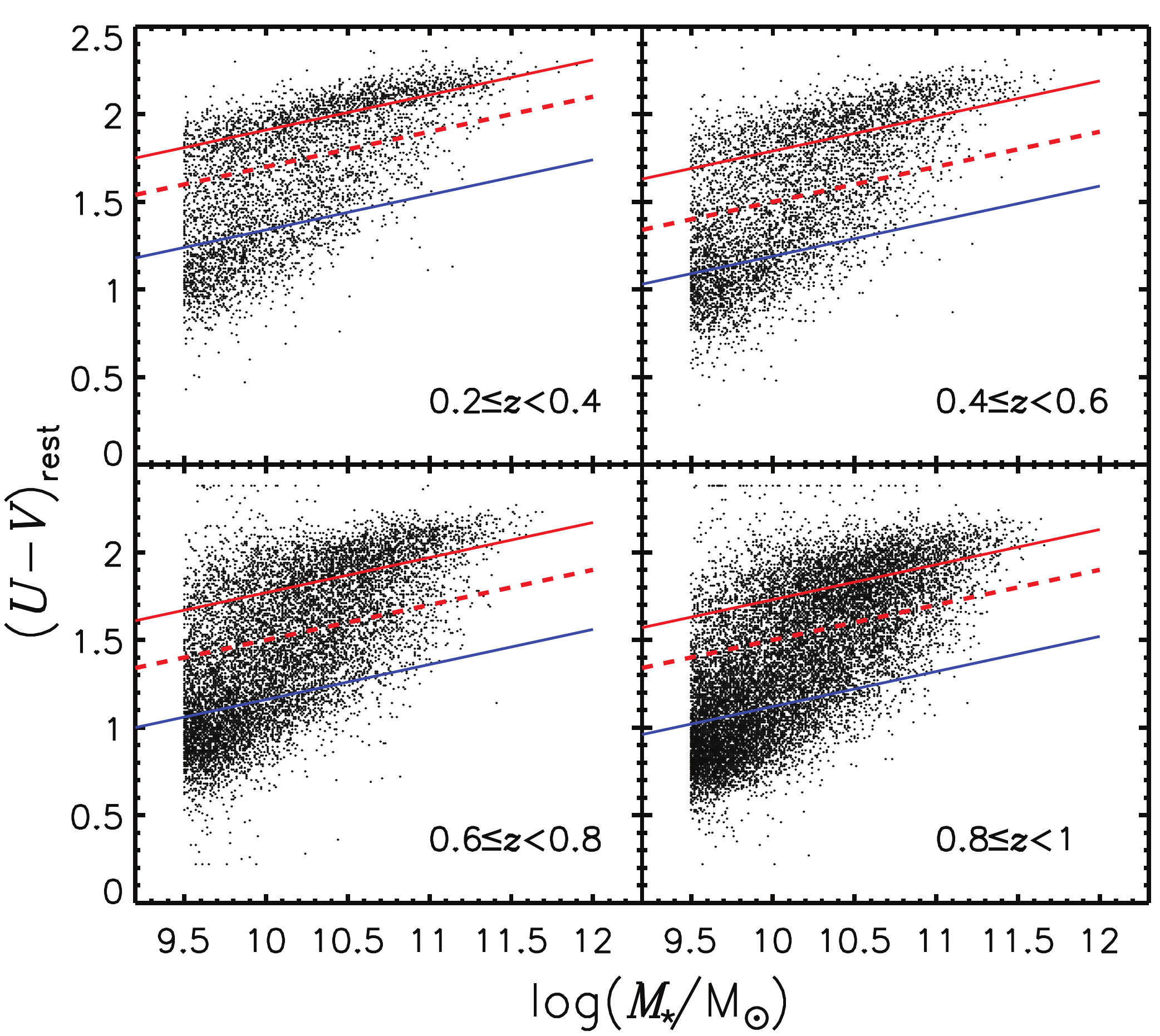}
\caption{The rest-frame $U-V$ as a function of stellar mass for  35\,076 galaxies with $\,M_\star\ge10^{9.5}\,\rm M_{\odot}$ and $0.2\leq z \leq 1$ in the COSMOS field. In each panel, the dashed line separates blue-cloud and red-sequence galaxies following \citet{2006A&A...453..869B}; and the red and blue solid lines are parallel to the dashed line and best fitting the two populations, respectively.}
\label{fig:figB}
\end{figure}

\begin{figure*}[]
\centering
\includegraphics[width=0.8\textwidth]{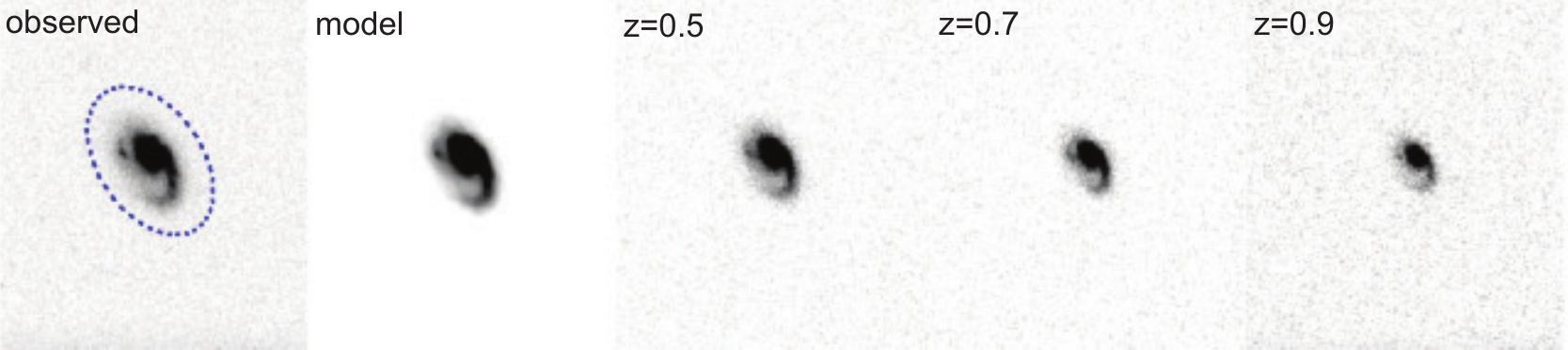}
\caption{Image comparison of one example galaxy at $z=0.37$ with the galaxy model  (i.e. the smoothed profile), and the same galaxy redshifted to $z\sim0.5$, 0.7 and 0.9 (from left to right).   A tidal arm clearly seen in the original science image (left) gradually shrinks with increasing redshift, denoting that the low-surface-brightness features are hardly detectable for high-$z$ galaxies. The blue dashed ellipse represents 1.5 times Petrosian radius of the target. The semi-major axis of such ellipse is 10.7\,kpc and the semi-minor axis is 6.8\,kpc.}
\label{fig:figC}
\end{figure*}

      The software tool FERENGI \citep[Full and Efficient Redshifting of Ensembles of Nearby Galaxy Images,][]{2008ApJS..175..105B} is used to do simulations. To redshift a model galaxy to high $z$,  one needs to correct for size downscaling, surface brightness dimming and band-shifting. The size degradation and brightness dimming can be directly calculated when the cosmology is given. Moreover, the intrinsic size evolution of galaxies since $z=1$ \citep{2014ApJ...788...28V} is considered in our simulation because galaxies are physically more compact in the past. Additionally, the evolution of stellar population in galaxies complicates the brightness estimate. Instead,  we determine the brightness of a redshifted galaxy from real galaxies at the high $z$ with the same stellar mass as the low-$z$ galaxy.  The counterparts at high $z$ are identified to have the same location in the stellar mass versus rest-frame $U-V$ color diagram.
%
Figure~\ref{fig:figB} shows the stellar mass versus rest-frame $U-V$ diagram of our sample of 35\,076 galaxies from the UltraVISTA catalog. The blue-cloud and red-sequence galaxies are separated by the dashed line in each of four even redshift bins over 0.2$\leq z \leq$1 following \citet{2006A&A...453..869B}.  In each panel of  Figure~\ref{fig:figB},  the red and blue solid lines, parallel to the dashed line,  describe the mass-color relations of two populations, respectively.  At a fixed stellar mass, the offset from one mass-color relation normalized by the distance of the relation from the dashed line marks the location of a galaxy in the diagram.  By doing so, we obtain locations of the selected 1178 galaxies used for simulations in the mass-color diagram of 0.2$\leq z <$0.4.   In each of the other three redshift bins ($z\sim0.3$, 0.5 and 0.9), we select ten nearest real galaxies centered at each of the 1178 locations and take the median $I$-band magnitude of the ten galaxies as the brightness of the model galaxy to be simulated.
In addition, photon noise  is added into the model images of simulated high-$z$ galaxies. The same sky background randomly selected from blank regions of the \textit{HST}/ACS $I$-band images of COSMOS are also combined with the model images. For each model galaxy, we create 16 background images and thus general 16 model images used for measuring $A_{\rm O}$ and $D_{\rm O}$.
Figure~\ref{fig:figC} presents the original science image of one example galaxy at low-$z$, in comparison with the smoothed profile and images of the same galaxy redshifted to $z\sim0.5$, 0.7 and 0.9.

\begin{figure*}[]
\centering
\includegraphics[width=0.8\textwidth]{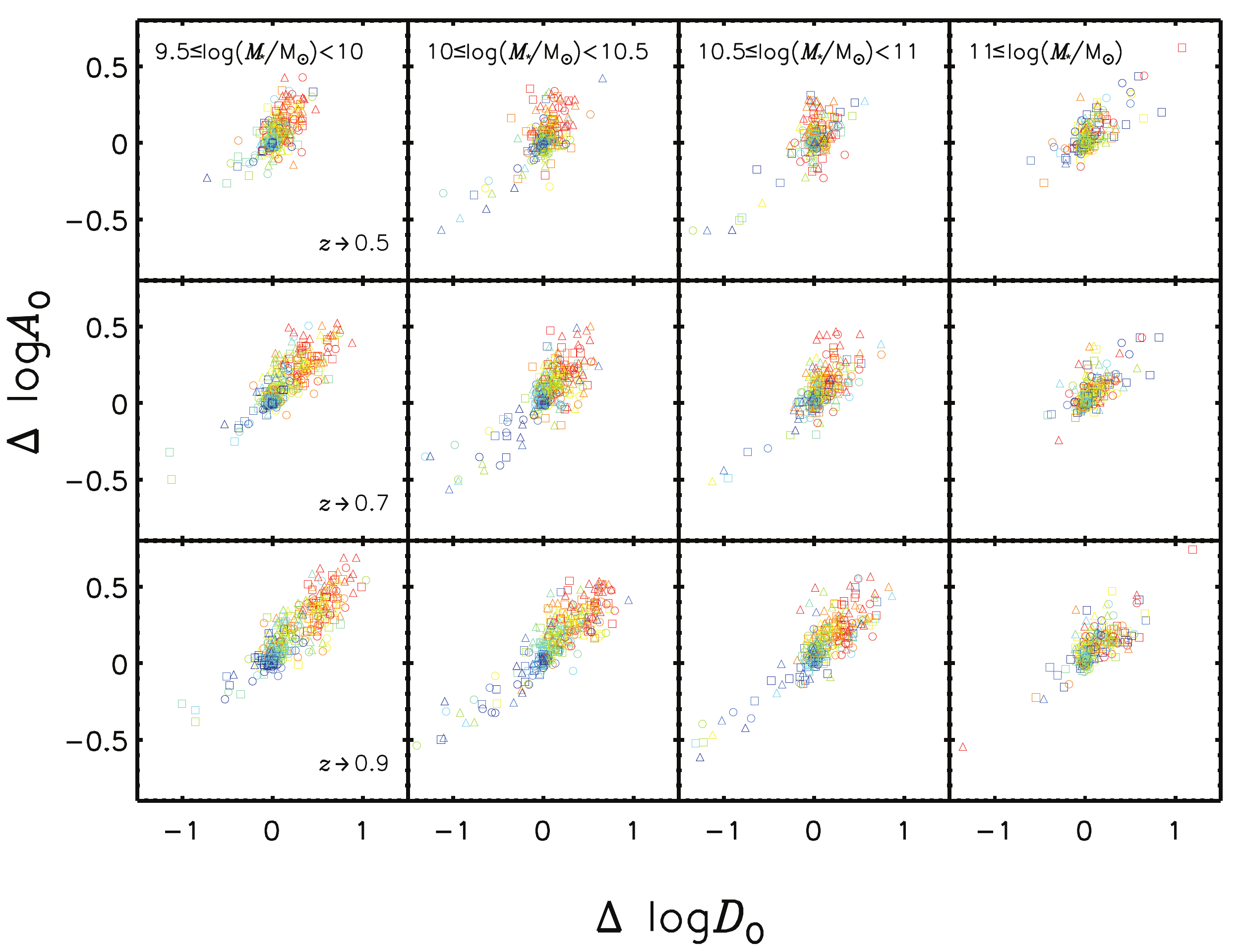}
\caption{Deviations of measured $A_{\rm O}$ and $D_{\rm O}$ for same model galaxies at $z\sim0.5$ (left), 0.7 (middle) and 0.9 (bottom). Mathematically, $\Delta \log A_{\rm O}=\log A_{\rm O}(z)-\log A_{\rm O}(\rm model)$ and $\Delta \log D_{\rm O}=\log D_{\rm O}(z)-\log D_{\rm O}(\rm model)$. Colors and symbols are same as in Figure~\ref{fig:figA}.}
\label{fig:figD}
\end{figure*}

       The same configurations for \textit{HST} images given in Section~\ref{sec:sec2} are adopted to measure $A_{\rm O}$ and $D_{\rm O}$ from the images of simulated high-$z$ galaxies with morphology known.  Background noise of course affects the detection of signals from a simulated galaxy, in particular in the outskirts with low-surface brightness. Each simulated galaxy image is combined with 16 random background images to estimate errors in measured $A_{\rm O}$ and $D_{\rm O}$ caused by the background noise. We find marginal variations among measured $A_{\rm O}$ and $D_{\rm O}$, suggesting that the measurements of the two structural parameters are reliable and show no dependence on the type of morphologies.
 The mean of 16 $A_{\rm O}$ ($D_{\rm O}$) measures is adopted for a simulated galaxy. Then we compare the measures of $A_{\rm O}$ and $D_{\rm O}$ at given redshifts with these derived from the (noise-free) model galaxy image.  Figure~\ref{fig:figD} demonstrates the changes of measured $A_{\rm O}$ and $D_{\rm O}$ when model galaxies with different stellar masses are located at $z\sim0.5$ (top), 0.7 (middle) and 0.9 (bottom). We can see that the changes of $A_{\rm O}$ and $D_{\rm O}$ are coupled with each other;  and then changes become larger at increasing redshift and for lower mass galaxies.  It is clear that the two parameters tend to be overestimated for galaxies with regular morphologies (intrinsically small in $A_{\rm O}$ and $D_{\rm O}$) and underestimated for those with disturbed morphologies (intrinsically large  in $A_{\rm O}$ and $D_{\rm O}$).  This is understandable because the relatively increasing contribution from noise easily increases $A_{\rm O}$ (and $D_{\rm O}$) for the regular galaxies when they become faint and less resolved at increasing redshift; however, the galaxies displaying irregular structures in their outskirts would appear more regular (measured $A_{\rm O}$ and $D_{\rm O}$ decreasing ) as the faint asymmetric features become gradually undetectable at increasing redshift.  And lower-mass galaxies are relatively more influenced  by these effects.
We conclude from Figure~\ref{fig:figD} that measures of $A_{\rm O}$ and $D_{\rm O}$ for high-$z$ galaxies may be systematically biased in the sense that  the spread of galaxies in both $A_{\rm O}$ and $D_{\rm O}$ would  shrink  to have more in the middle at increasing redshift. This suggests that part of high-$z$ galaxies with disturbed morphologies would be missed in our $A_{\rm O}-D_{\rm O}$ selection and merging galaxies with faint tidal tails are barely detectable at high $z$. Bearing this in mind, we try to correct for the incompleteness to estimate the merger rate in Section~\ref{sec:completeness}.

\section{Merging Galaxies with Long Tidal Tails}

\subsection{Identification of Merging Galaxies with Tidal Tails}\label{sec:ideT}

\begin{figure*}[]
  \centering
    \includegraphics[width=0.49\textwidth]{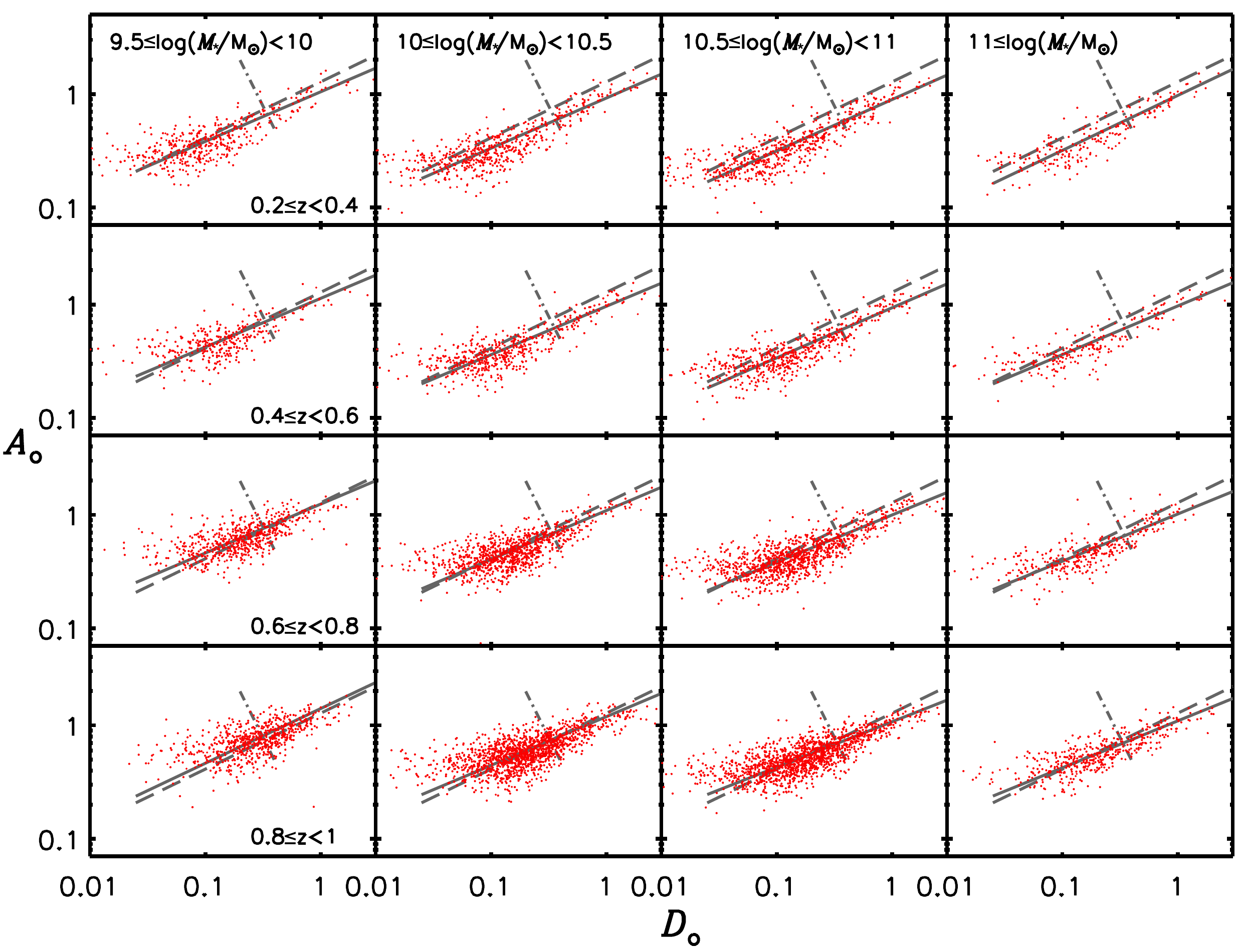}
    \includegraphics[width=0.49\textwidth]{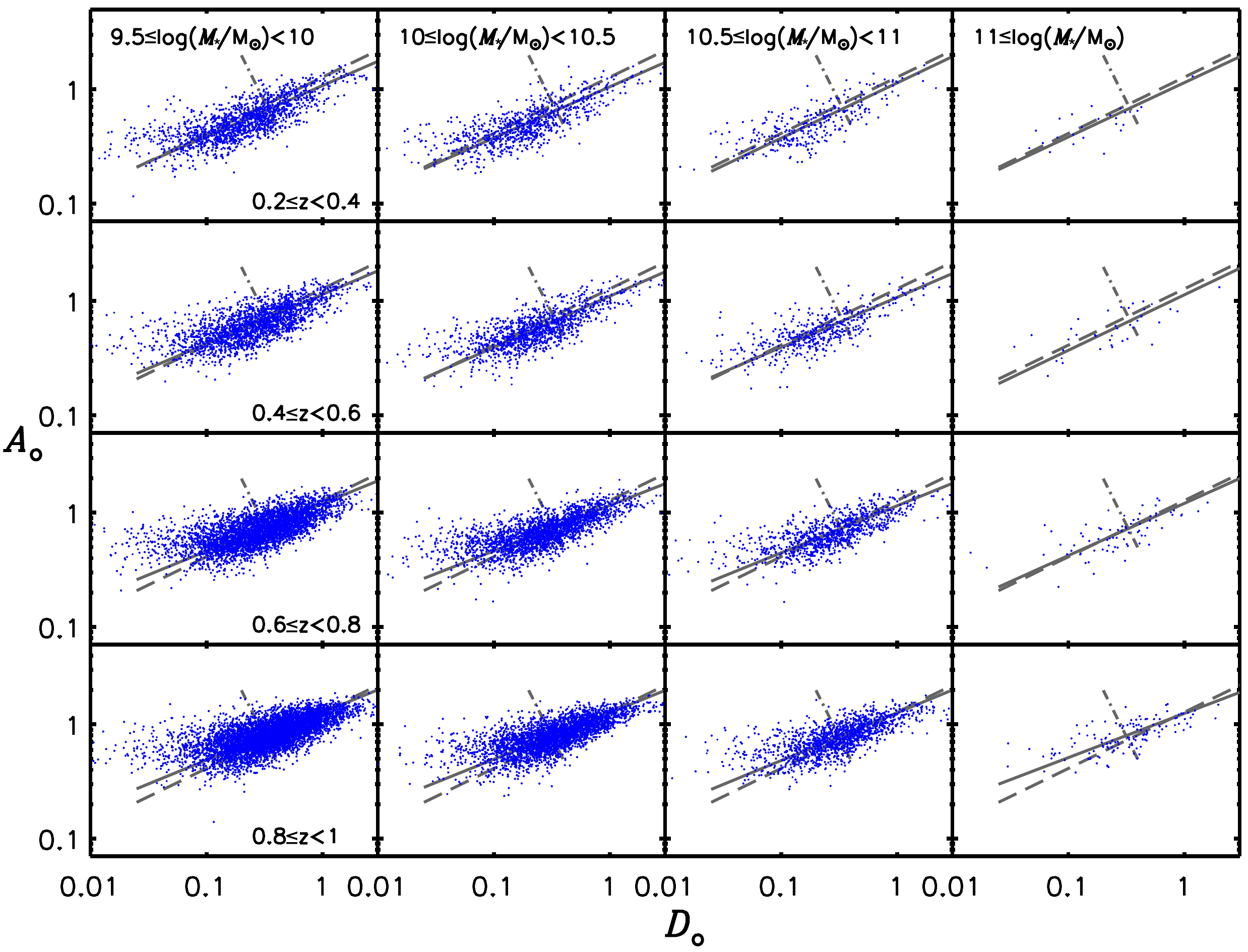}
  \caption{The $A_{\rm O}-D_{\rm O}$ diagram of 35\,076 galaxies from COSMOS split into 4 mass bins and 4 redshift bins. The left panels show  quiescent galaxies (red) while the right panels show star-forming galaxies (blue). The solid line in each bin is the best fit to the data points. The dashed line represents the best fit of galaxies in the CDFS field from \citet{2014ApJ...787..130W}. The dashed-dotted line gives the selection cut for galaxies with tidal tails (see the Appendix for more details).}
\label{fig:figFG}
\end{figure*}

We measure $A_{\rm O}$ and $D_{\rm O}$ from \textit{HST}/ACS $I$-band images for the complete sample of 35\,076 galaxies with $\log (M_\star/M_\odot)\ge 9.5$ and $0.2\le z\le1$ in the COSMOS field.  The sample galaxies are divided into star-forming and quiescent galaxies using the $UVJ$ method \citep{2009ApJ...691.1879W}.
Figure~\ref{fig:figFG} presents the measurement results of $A_{\rm O}$ and $D_{\rm O}$ for the star-forming and quiescent galaxies split into 4 bins in stellar mass and 4 bins in redshift.   The best-fit solid lines to subsamples of galaxies have similar slopes,  indicating that the correlation between $A_{\rm O}$ and $D_{\rm O}$ is nearly uniform. As pointed out in Section~\ref{sec:simulation},  galaxies at $z>0.4$ appear to be more concentrated in the $A_{\rm O}-D_{\rm O}$ space, especially for less massive galaxies. Still, we find that star forming galaxies have larger $A_{\rm O}$ and $D_{\rm O}$ than quiescent galaxies, which are usually smooth in morphology.
Although some faint asymmetric features in galaxy outskirts are undetectable at high $z$, leading to the underestimate of $A_{\rm O}$ and $D_{\rm O}$, these with measured $A_{\rm O}$ and $D_{\rm O}$ sufficiently high should have  asymmetric structures detectable.
We note that foreground or background sources may project next to target galaxies and  result in false asymmetric structures. Such cases are removed in our visual examination.

We aim to estimate the rate of merging galaxies traced by tidal tails as a function of redshift.
\citet{2014ApJ...787..130W} investigated the selection criterion for merging galaxies using the $A_{\rm O}-D_{\rm O}$ method, yielding that galaxies with long tidal tails satisfy $\log(A_{\rm O}) = -2\,\log(D_{\rm O})-1.1$.  We adopt the criterion in our analysis, shown as the dashed-dotted lines in Figure~\ref{fig:figFG}, and select 13\,227 of 35\,076 galaxies to be candidates for merging galaxies.
The next step is to visually check morphologies of the 13\,227 galaxies to identify merging galaxies with long tidal tails.  We point out that the 13\,227 galaxies must contain mergers with other morphologies \citep[e.g.,][]{2014AJ....148..137L} and a detailed analysis of these galaxies will be presented in a future work.

\begin{figure*}[]
\centering
\includegraphics[width=0.95\textwidth]{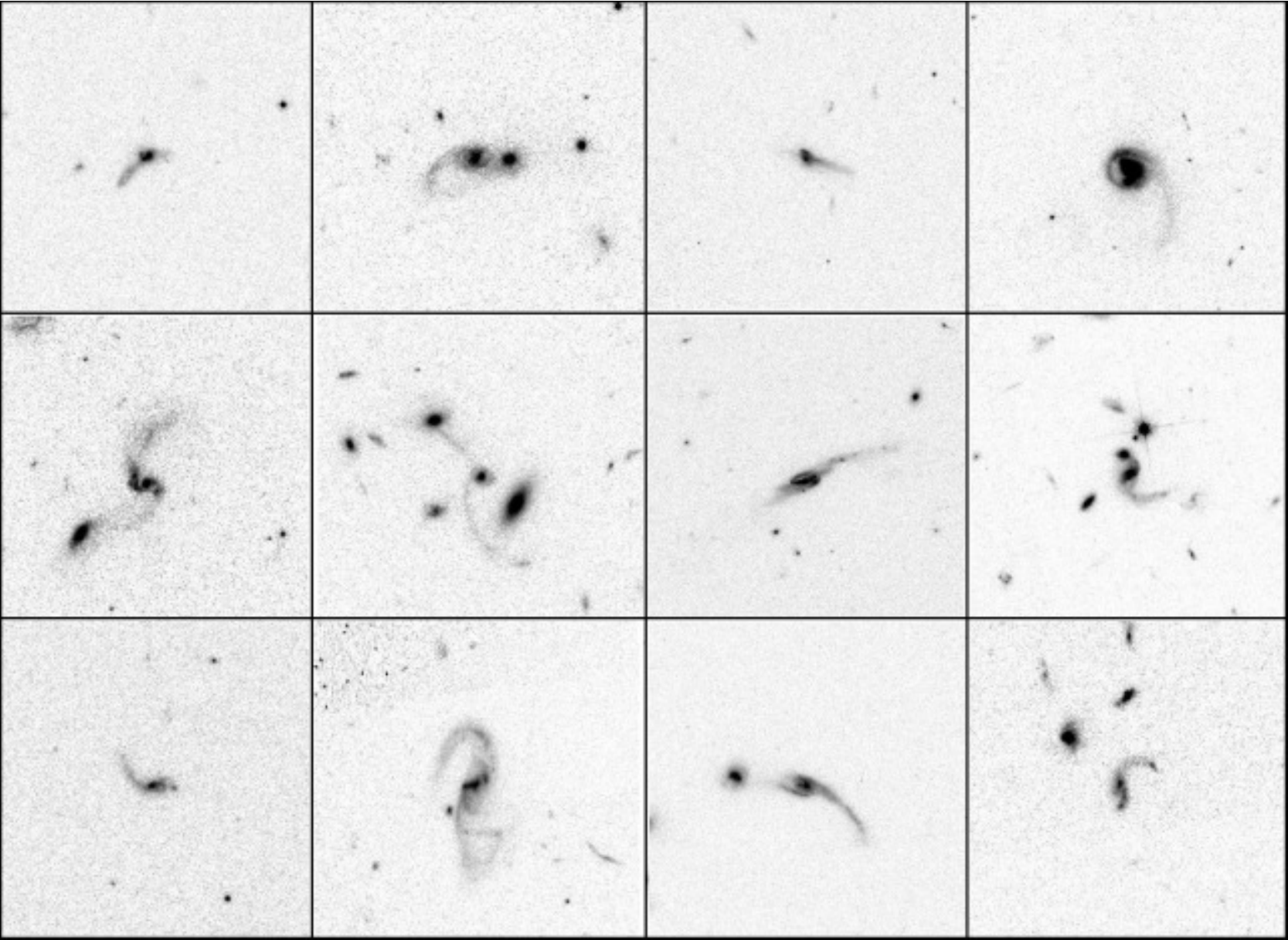}
\caption{Examples of mergers with long tidal tails from COSMOS. These image stamps are given in a size of 120$\times$120\,kpc. All of the 461 stamps are available in the online material.}
\label{fig:figH1}
\end{figure*}
\begin{figure*}[]
\centering
\includegraphics[width=0.95\textwidth]{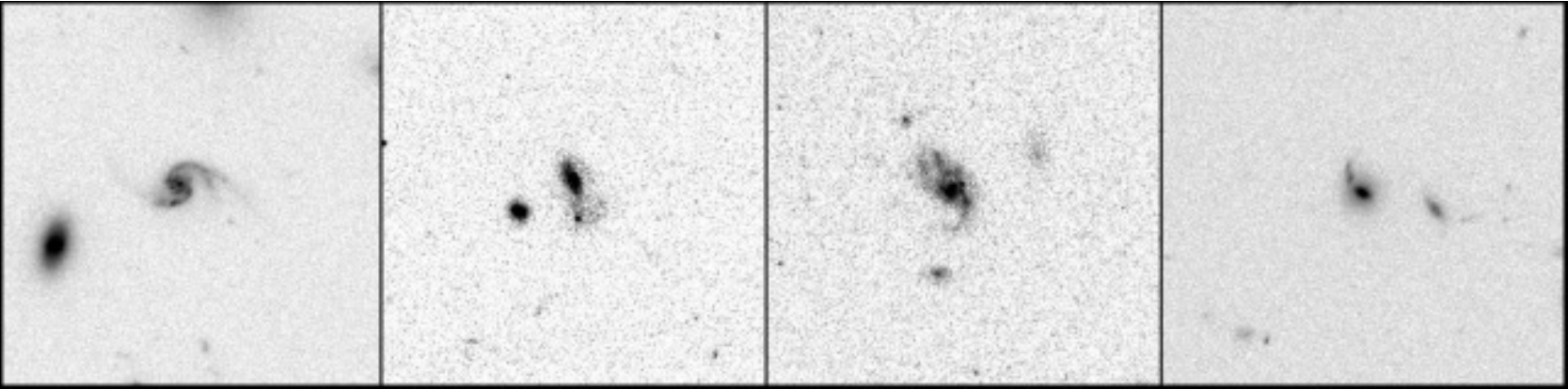}
\caption{Examples of galaxies with tail-like structures (spiral arms, accreting bridges from companions, short tidal tails and debris related with early-type galaxies) but not long tidal tails.  These image stamps are given in a size of 120$\times$120\,kpc. }
\label{fig:figH2}
\end{figure*}

Due to the projection effect, a tidal tail may appear to be shorter. On the other hand, tidal tails rapidly grow during a merging process and the length of the tails is dependent on the stage of the merging process.  In our examination, the length of a tail is measured from the edge of the host galaxy to the tip end of the tail. The edge is approximately determined by the isophotal ellipse of the host galaxy when the tails are masked.  Because short tidal tails are often indistinguishable from spiral arms or tidal arms (tidally-disturbed spiral arms), we rather focus on long (or extended) tidal tails. Here the long tails refer to these with length equal to or longer than the diameter of their host galaxies. The visual identification of tidal tails is independently carried out by Z.~Z.~W and X.~Z.~Z. Totally 461 galaxies are confirmed by the two viewers with long tidal tails. Figure~\ref{fig:figH1} illustrates 12 examples from the 461 galaxies. They are apparently merging systems with long tidal tails.  We also show four examples of galaxies exhibiting tail-like structures in Figure~\ref{fig:figH2}. These structures are either spiral arms, accreting bridges from companions, or short tidal tails, and not included into our sample of long-tidal-tail systems.

In total 461 galaxies are identified to have long tidal tails from 13\,227 merger candidates selected using the $A_{\rm O}-D_{\rm O}$ technique  from the parent sample of 35\,076 galaxies with  $\log (M_\star/M_\odot)\ge 9.5$ and $0.2\le z\le1$ in the COSMOS field. The photometric catalogs of the 461 galaxies and their image stamps  are electronically available online. We note that if two galaxies in a merging system are both resolved in the ground-based $K_{\rm s}$-band catalog, they are counted as one merger event. The stellar mass of such a system is the sum of two components. We also find some of them to contain star clusters and tidal dwarf galaxy (TDG) candidates in their tidal tails.  We refer massive compact clumps found in tidal tails to be TDG candidates. TDGs are self-gravitating, dark-matter-free systems formed by gravitational instability of pre-enriched gas clouds in tidal tails with size and mass comparable to classical dwarf galaxies \citep{2001A&A...378...51B,2003A&A...397..545W,2004A&A...427..803D,2006A&A...456..481B,2007MNRAS.375..805W,2010AdAst2010E...1B,2012Ap&SS.337..729W}. Due to the lack of multiband high-resolution imaging data, we are unable to extract spectral energy distributions for the TDGs candidates and estimate stellar mass and properties of stellar population.  Still, we provide a sample of 181 TDG candidates at intermediate redshifts,  which can be used for further exploration.

\begin{table*}[]
\begin{center}
\caption{\label{tab:tab0} Catalog of Merging Galaxies with Tidal Tails}
\begin{tabular}{ccccccccccc}
\hline
\noalign{\vskip 0.05cm}
\hline
ID & RA & Dec. & $z$  & $\log (M_\star/M_\odot)$ &  $M_{\rm I} $  & $M_{\rm K_s}$ &  $U-V$  & $V-J$  &  $D_{\rm O}$  & $A_{\rm O}$ \\
 & (h m s) & (d m s) &   &  &  (mag)  & (mag) &   &   &  &  \\
\hline
         218 &  10 01 38.323   & 01 38 27.50 & 0.94 & 10.2 & 23.0 & 21.1 & 1.52 & 1.05 & 1.27 & 1.20 \\ 
         708 &  10 01 34.615   & 01 39 57.87 & 0.60 & 9.66 & 20.5 & 19.8 & 0.40 & 0.51 & 1.40 & 1.31 \\ 
         737 &  10 02 32.352   & 01 39 59.63 & 0.85 & 10.1 & 21.6 & 20.5 & 0.97 & 0.81 & 0.57 & 1.04 \\ 
         939 &  10 02 21.446   & 01 40 21.89 & 1.00 & 10.4 & 21.6 & 20.1 & 0.97 & 0.79 & 0.39 & 0.86 \\ 
...&...&...&...&...&...&...&...&...&...&...\\
\hline
\end{tabular}
\tablecomments{This table is available in its entirety in the online material. A portion is shown here for guidance regarding its form and content.}
\end{center}
\end{table*}

\begin{figure}[]
\centering
\includegraphics[width=0.45\textwidth]{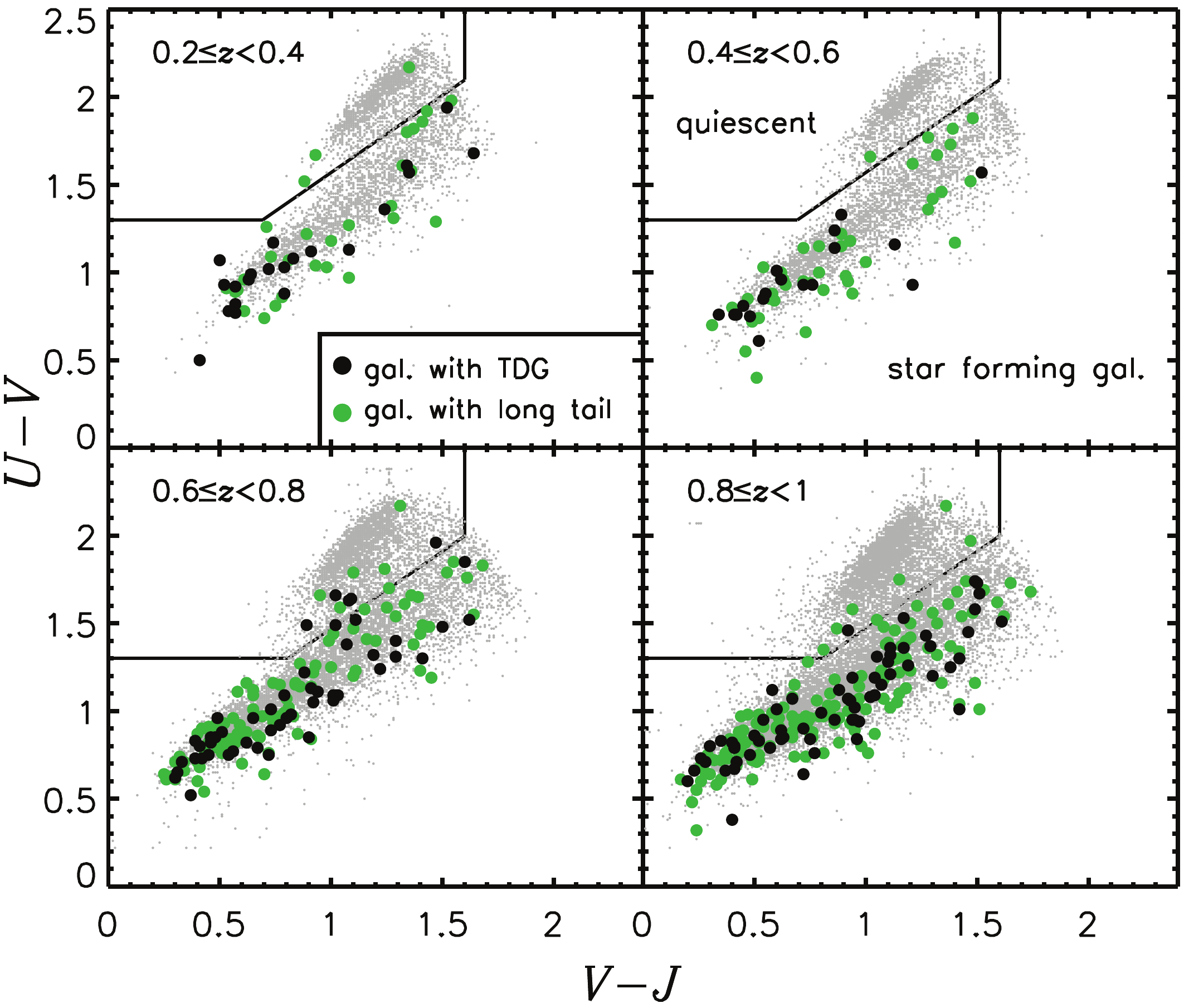}
\caption{The $U-V$ Vs. $V-J$ diagram of 461 merging galaxies with long tidal tails (solid circles) in comparison with the parent sample of 35\,076 galaxies (gray points).  The black (green) circles represent  mergers containing (or not) tidal dwarf galaxy candidates. The solid lines show the adopted division between the star-forming and quiescent galaxies in each redshift bin.}
\label{fig:figI}
\end{figure}

The selected 461 merging galaxies with long tidal-tails are compared with the parent sample of galaxies in the  $U-V$ versus $V-J$ color diagram, as shown in Figure~\ref{fig:figI}. The grey data points represent quiescent and star-forming galaxies clearly separated into two regions in this plot. The 461 tidal-tailed galaxies with (without) TDG candidates are shown in black (green) solid circles. The adopted selection criteria for quiescent galaxies are derived from \citet{2009ApJ...691.1879W}. We can see that the solid circles widely spread in the region of star-forming galaxies, suggesting that the tidal-tail-selected merging galaxies originate from this population. Only twenty-four tidal-tailed systems fall into the region of quiescent galaxies, indicating that they are likely to be in the late stage of merging with star formation quenched already.
This result that tidal-tailed galaxies belonged to the star-forming galaxy population is consistent with  our understanding of the tidal-tail creation in merger events involving disk galaxies that usually are forming stars. In contrast, mergers between two early-type galaxies often display diffuse tidal debris with no or only little star formation \citep[e.g.,][]{2007ApJ...663..734E}. The diffused structures are generally arc-like or plume-like, having similar colors as their host galaxies. Although such tidal structures may occasionally  appear as tail-like in a special projection angle, mergers between early-type galaxies are still ignorable in a statistic sense \citep{2013LNP...861..327D}.


\begin{figure}[]
\centering
\includegraphics[width=0.45\textwidth]{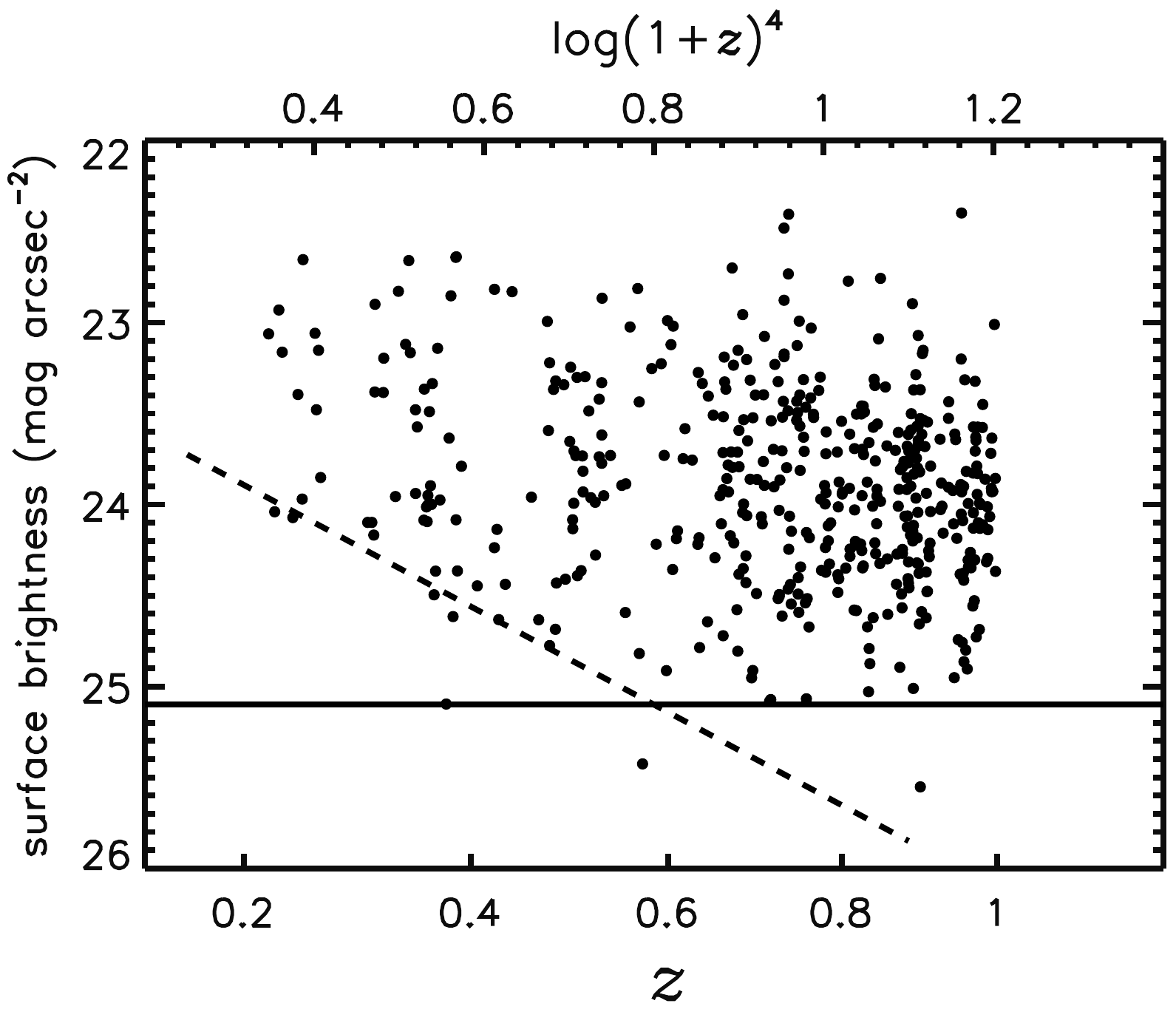}
\caption{ The $I$-band surface brightness of 461 tidal tails. The solid line gives the 3\,$\sigma$ surface brightness limit. The dashed line shows the cosmic dimming as $\sim (1+z)^4$ for a tidal tail with intrinsic surface brightness of 23.1\,mag\,arcsec$^{-2}$ in optical bands.}
\label{fig:figSB}
\end{figure}

Moreover, the $I$-band surface brightness distribution of identified tidal tails is given in Figure~\ref{fig:figSB}. The measurements are performed using the task \textit{Polyphot} in IRAF. The surface brightness of tidal tails is within $22-24$\,mag\,arcsec$^{-2}$ at $z\sim0.2$ down to $23-25$\,mag\,arcsec$^{-2}$ at $z\sim1$ due to the surface brightness dimming effect. Only a few tidal tails are fainter than the $3\,\sigma$ surface brightness limit of 25.1\,mag\,arcsec$^{-2}$. The surface brightness limit of the $I$-band imaging data is determined using aperture photometry in a blank sky area. The photometry is performed and repeated using an aperture of radius $=5$ pixels ($\sim 1.5$ PSF size). As shown in Figure~\ref{fig:figSB}, nearly all the identified tidal tails have intrinsic surface brightness $<$ 23.1\,mag\,arcsec$^{-2}$ in optical bands. Comparing with the detection limit of the $I$-band data, our selection of tidal tails becomes incomplete at high redshift. The incompleteness of tidal tail selection will be estimated through simulations (see Section~\ref{sec:completeness}) and corrected in merger rate calculation.

\subsection{Estimate of Completeness }\label{sec:completeness}

       It is important to address the completeness of mergers with tidal tails selected by the $A_{\rm O}-D_{\rm O}$ technique. Tidal tails of mergers at high $z$ may be spatially resolved, but undetectable if they appear to be too fainter due to cosmological surface brightness dimming. From simulations presented in Section~\ref{sec:simulation}, we estimate the fractions of tailed galaxies that are not picked up by the $A_{\rm O}-D_{\rm O}$ selection together with visual examination when the tailed galaxies at $0.2<z<0.4$ are redshifted to $z\sim0.5, 0.7$ and 0.9. Of 60 merging galaxies with long tidal tails, 47 (78\%), 43 (72\%) and 37 (62\%) are recovered as the tailed ones at $z\sim0.5, 0.7$ and 0.9, respectively. We use these fractions to correct for the incompleteness in estimating occurrence rate of mergers with long tidal tails. It is worth noting that the estimation of completeness is based on two assumptions: (1) The intrinsic surface brightness distribution of tidal tails is not changed significantly across $0.2<z<1$ (see Section~\ref{sec:ideT}); (2) The completeness of tidal tails at the lowest redshift bin $z\sim0.3$ is 100\% and the fractions estimated at higher redshift are relative to the completeness at $z\sim0.3$.

\section{Galaxy Merger Rate}\label{sec:fraction}

Long tidal tails are most likely produced in major mergers of two parent galaxies and at least one disk galaxy is involved \citep{2013LNP...861..327D}. Stars in the parent disk galaxy had ordered motions (i.e., following a common rotational velocity pattern) around the spin axis. Tidal force from companion may change the dynamics of the stars in the same way and thus form prominent tidal features like tails after the first pericenter passage of galaxy encounter. However, stars in pressure supported galaxies (e.g., ellipticals) fail to follow a uniform motion when external tidal force is applied. Therefore, only diffused tidal debris formed in the merger between two early-type galaxies.
 
As explored by numerical simulations \citep{1992ApJ...393..484B,1993ApJ...417..502H,1996ApJ...464..641M,2013LNP...861..327D,2016MNRAS.455.1957B}, characterization of the tidal tails is strongly correlated with the orbital parameters (e.g. spin and orbital angular momentum) of the mergers. In the case of a planar prograde encounter, the spin and the orbital motion are coupled (zero-inclination) and the strongest tidal tail can be created. As the disk inclination angle increases, the tails tend to be more warped. In the case of planar retrograde encounter, only weak tidal features can be formed. The galactic disks often exhibit short-lived ripple-like features instead of a tidal bridge or tail. Meanwhile, highly-warped and diffused structures can be formed in highly-inclined retrograde encounter. Only at extreme conditions, the polar encounter could create a tail-like broad and straight tidal feature. In this work, mergers are identified based on their long tidal tails, indicating that the mergers selected are most likely undergoing prograde encounters and bias against retrograde encounters.


\subsection{Galaxy Merger Fraction}

\begin{figure}[]
\centering
\includegraphics[width=0.45\textwidth]{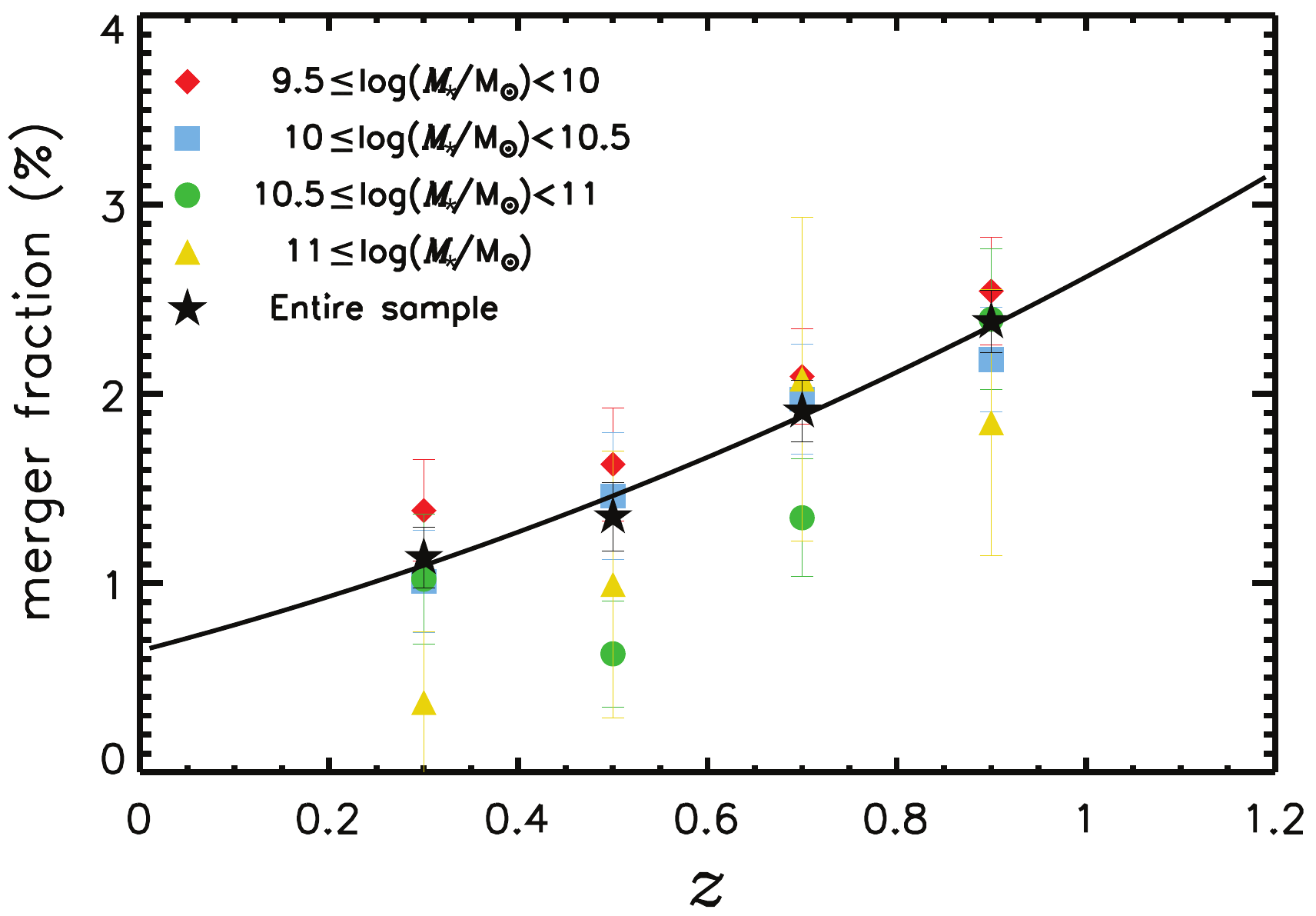}
\caption{Fraction of merging galaxies with tidal tails as a function of redshift. The colored points represent the galaxies with different stellar mass while the black points give the fraction of the tailed mergers over entire mass range.  The best-fit power-law function to the black data points are given as $(1+z)^m$ where $m=2.0\pm0.4$. }
\label{fig:figJ}
\end{figure}

Using the sample of 461 merging galaxies with long tidal tails, we estimate the fraction of such galaxies among the galaxies of similar stellar mass and redshift. Using the completeness corrections presented in Section~\ref{sec:completeness}, we estimate the merger fraction of galaxies with stellar mass $\ge 10^{9.5}\,M_\odot$. The numbers of tailed galaxies in different stellar mass and redshift bins are listed in Table~\ref{tab:tab1}. The merger fractions that are corrected for surface brightness incompleteness are also shown in Figure~\ref{fig:figJ}. We find that  merger fraction appears to be higher at higher redshifts, consistent with the theoretical predictions from the hierarchical cosmology \citep{2010ApJ...724..915H}. The evolution of merger fraction with redshift can be fitted by a power-law function in the form of $f =f_0\,(1+z)^m$. We adopt the Monte Carlo method to take the uncertainties of data points into account and derive the the best-fit parameters of the function from 3\,000 fits using the least squares method, yielding $m=2.0\pm0.4$ and $f_0=0.64\pm0.13$ percent.
Moreover, our results reveal that merger fraction is dependent on galaxy stellar mass, as shown in Figure \ref{fig:figJ}, in the sense that  less massive galaxies exhibit a higher fraction of mergers than massive galaxies out to  $z=1$, although uncertainties are relatively large.

\begin{table*}[]
\begin{center}\scriptsize
\caption{\label{tab:tab1} Numbers of Mergers and Galaxies}
\begin{tabular}{cccccccc}
\hline
\hline
 $z$ & \multicolumn{4}{c}{$\log (M_\star/M_\odot)$} & $N_{\rm tot}$  &Completeness &Corrected Merger\\
\cline{2-5} 
 & $9.5-10$ & $10-10.5$ & $10.5-11$ & $>11$ & &  & Fraction [$\%$]\\
\hline
0.2 $-$ 0.4   &  27/1951    &  14/1387   &  9/881      &  1/270  &  51/4489     & 1      & 1.1$\pm$0.16\\
0.4 $-$ 0.6   &  30/2353    &  19/1662   &  5/1019    &  2/257  &  56/5291     & 0.78 & 1.4$\pm$0.18\\
0.6 $-$ 0.8   &  69/4606    &  47/3328   &  19/1972  &  6/403  &  141/10309 & 0.72 & 1.9$\pm$0.16\\
0.8 $-$ 1.0   &  101/6618  &   63/4815  &  42/2923  &  7/631  &  213/14987 & 0.62 & 2.4$\pm$0.17\\
\hline
\end{tabular}
{\tablecomments{Numbers of merging galaxies with long tidal tails and parent galaxies in stellar mass and redshift bins. The errors are inferred from poisson uncertainties of the numbers of tailed galaxies and parent galaxies through error propagation. Noting that the sample completeness is 95\% for galaxies in the lowest mass bin at $z\sim0.9$. Such uncertainty is also considered in error estimation.}}
\end{center}
\end{table*}

\subsection{Galaxy Merger Rate}

Galaxy merger rate is usually described in two ways. One is the fractional merger rate $\Re$ that traces the number of mergers that one galaxy would undergo in unit time, given by
\begin{equation} \label{equ:eq4}
\Re =f_{\rm m}/ T_{\rm m}.
\end{equation}
The other is the merger rate $\Gamma$ in units of number of mergers per unit time per unit co-moving volume:
\begin{equation} \label{equ:eq5}
\Gamma =n_{\rm gal} \ast f_{\rm m}/ T_{\rm m}.
\end{equation}
Here $f_{\rm m}$ is galaxy merger fraction, $T_{\rm m}$ is the average detection timescale during which the merger is selected by a specific technique, and $n_{\rm gal}$ is the number density of galaxies in the co-moving volume at a given redshift. Although the merger rate $\Gamma$ provides a direct measure of merging events in the framework of galaxy formation and evolution, it suffers from large uncertainties as $n_{\rm gal}$ varies from field to field due to the cosmic variance. The fractional merger rate $\Re$ was often presented in the literature.  We also estimate the fractional merger rate using our sample of tailed galaxies.

The detection timescale $T_{\rm m}$  is a key parameter for determining merger rate of galaxies. In this work, major mergers are traced by long tidal tails and the duration for displaying long tidal tails in merger processes is thus used as $T_{\rm m}$. The duration can be determined from merger simulations. From the simulations by \citet{2006ApJ...638..686C}, we carefully investigate the snapshots of mergers between equal-mass disk galaxies. Long tidal tails are visible when the host galaxies reach their maximal separation (0.68\,Gyr) and can be still seen at 1.2\,Gyr when disk progenitors have merged as a single system. However, they become invisible at 1.33\,Gyr, suggesting that duration for displaying long tidal tails is around 0.5\,Gyr or longer. Meanwhile, some simulations gave durations for long tidal tails, ranging from about 0.25\,Gyr \citep{2006A&A...456..481B,2011ApJ...730....4B} to 0.77\,Gyr \citep{2014A&A...566A..97J}. Finally, we adopt 0.5$\pm$0.25\,Gyr to be the duration for displaying long tidal tails in major mergers. It is worth noting that identifying long tidal tails in snapshots is dependent on the viewing angle. An averaged duration for long tidal tail appearance over viewing angles would improve the estimate of duration.

 \citet{2009ApJ...697.1971J} adopted $T_{\rm m}\sim0.5-0.8\,\rm Gyr$ for visually selected gas-rich mergers. The adopted timescale of tailed mergers in this work is the lower limit of that in \citet{2009ApJ...697.1971J} for that tidal tails only account for a certain merging stage. In addition, a longer timescale (0.8\,Gyr) is used in B10, aimed at accounting for longer merger stage from close-pair with tidal bridges to the later coalescence stage exhibiting long tidal tails.

\begin{figure}[]
\centering
\includegraphics[width=0.45\textwidth]{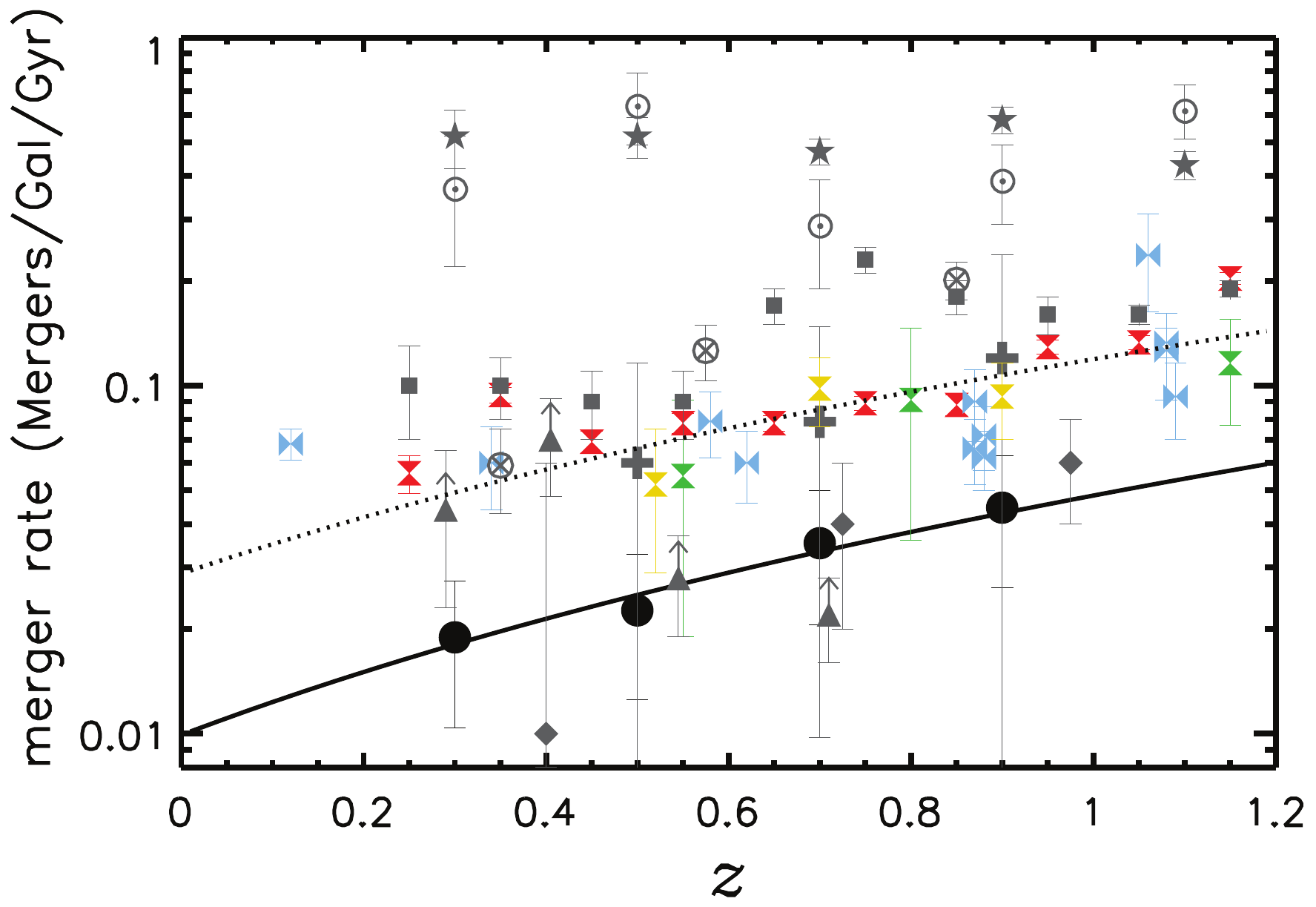}
\caption{The major merger rate of galaxies with $\log(M_\star/M_{\odot})\ge10$ as a function of redshift in comparison with representative results from the literature. All the grey symbols refer to merger rates based on disturbed morphologies while the colored symbols represent merger rates based on close pairs. Our results are given in black solid circles best-fitted by $\propto (1+z)^{2.3\pm1.4}$. $Red\,\,hourglass$: \citet{2007ApJS..172..320K}; $Blue\,\,hourglass$: \citet{2008ApJ...681..232L}; $Yellow\,\,hourglass$: \citet{2009A&A...498..379D}; $Green\,\,hourglass$: \citet{2009ApJ...697.1369B}; $Grey\,\,solar$: \citet{2008ApJ...672..177L}; $Grey\,\,triangle$: \citet{2009ApJ...697.1971J}; $Grey\,\,star$: \citet{2009ApJ...697.1764S}; $Grey\,\,diamond$: \citet{2009A&A...501..505L}; $Grey\,\,square$: \citet{2009MNRAS.394.1956C}; $Grey\,\,plus$: B10. $Grey\,\,circle\,\,with\,\,cross$: \citet{2014AJ....148..137L}. Merger rates in the literature are derived from \citet{2011ApJ...742..103L} if available. The dotted line represents the best fitting line to all the measurements on close galaxy pairs.}
\label{fig:figK}
\end{figure}

Merger rate is dependent on galaxy stellar mass  \citep{2011ApJ...742..103L}. We further focus on merger rate among galaxies with $\log(M_\star/M_{\odot})\ge10$ and compare our results with those based on the samples selected by the same mass cut. Figure~\ref{fig:figK} shows our results of fractional merger rate as a  function of redshift, in comparison with representative results from previous works described in Section~\ref{sec:dis}. The major merger rate we derived evolves mildly with redshift, and is best described by {\rm $R_0 (1+z)^{2.3\pm 1.4}$\,Gyr$^{-1}$ where $R_0=0.01\pm0.007$}. We notice that our data points are generally lower than the merger rates derived from galaxy pairs  or disturbed morphologies. This systematic offset can be explained by the merger identification bias. Our sample of major mergers traced by long tidal tails is biased against mergers without tidal tails such as mergers between two early-type galaxies.  Therefore our  major merger rate accounts for only a fraction of merger rate in each redshift bin.  The discrepancy of our measurements from the others are partly due to the sample selection and cosmic variance. We come back to discuss this issue in next section.

       We compute the number of expected mergers for a local galaxy undergoing over a given cosmic period using
\begin{equation}
N = \int_{t_{1}}^{t_{2}} \frac{1}{\Gamma(z)} dt = \int_{z_{1}}^{z_{2}} \frac{1}{\Gamma(z)} \frac{t_{H}}{(1+z)} \frac {dz}{E(z)},
\end{equation}
where $\Gamma(z)$ is the duration that a galaxy exists before undergoing a merger as a function of redshift, $t_{H}$ is the Hubble time, and  $E(z) = [\Omega_{\rm M}(1+z)^{3}+\Omega_{k}(1+z)^2+\Omega_{\Lambda}]^{-1/2} = H^{-1}(z)$ \citep{1999astro.ph..5116H,2009MNRAS.394.1956C}. We derive that a galaxy would undergo about 0.3 wet major mergers since $z=1$ to the present day from our measurements or 2.9 wet mergers when the merger rate is extrapolated to $z=2$.


\section{Discussion}\label{sec:dis}

       Tidal tails are solid indicator for major merger systems in which at least one disk galaxy is involved. The mergers of two early-type galaxies hardly generate such features \citep{2013LNP...861..327D}. The merger rate from this work accounts only for merging galaxies with long tidal tails, corresponding to a special type among a variety of mergers classified by morphology (and merging stage). Then the question is what fraction of overall galaxy major mergers display long tidal tails?   We discuss the representative measurements of galaxy merger rate from the literature and our measurement respectively, and try to draw an answer to this question.

    \subsection{Measurements of Merger Rate from Previous Works}\label{sec:sum}

Many studies have been contributed to the measurement of merger fraction of galaxies using a variety of selection techniques, incompleteness corrections and parent galaxy samples. An agreement has been reached between these studies, giving that galaxy merger fraction globally increases with redshift.  Still, significant discrepancy can be seen between different measurements. 

In Figure~\ref{fig:figK} we collect representative measurements of merger rate corrected to roughly the same mass selection $\log(M_\star/M_{\odot})\ge 10$ and ensure the merger rates are measured approximately from the same galaxy population. It is worth noting that evolving luminosity cuts are performed for sample selection in the literature \citep[e.g., $M_V<-19.8-1.0z$][]{2007ApJS..172..320K,2008ApJ...681..232L,2009ApJ...697.1764S}. The luminosity cuts are based on a model of passive luminosity evolution (PLE) of galaxies. \citet{2011ApJ...742..103L} argued that such cuts may select similar galaxy population as those derived through a fixed stellar-mass cut $M_\star\ge10^{10}\,M_{\odot}$. We thus include their results in Figure~\ref{fig:figK} for comparison. We caution that major merger rates from \citet{2009ApJ...697.1971J} and \citet{2014AJ....148..137L} in the Figure \ref{fig:figK} are based on a sample of galaxy with $M_\star\ge 2.5\times10^{10}\,M_{\odot}$ and $M_\star\ge 10^{10.6}\,M_{\odot}$. We caution that their selection cuts are higher than ours, although we include these results for comparison.

We can see that merger rates of close pairs are consistent with each other within the uncertainties, and the global merger rate can be best fitted by $0.03\pm0.002(1+z)^{2.0\pm0.1}$. This indicates that a galaxy would undergo 0.8 major (wet$+$dry) mergers since $z=1$, about 2.5 times higher than our measurement on mergers with tidal tails. Note that the merger rates of close pairs are converted from pair fractions ($f_{\rm pair}$) instead of fractions of galaxies in pairs ($N_c$)  \citep{2011ApJ...742..103L}, allowing comparison with merger rates from morphology studies. For this reason, two galaxies linked by a tidal bridge are counted as one merger event in this work. On the other hand, merger rate derived using morphological techniques are usually higher than those derived from close pairs. This is because the morphological selection for major mergers may be contaminated by minor mergers. Furthermore, galaxy pairs are early-stage mergers and thus correspond to progenitors of disturbed merging galaxies. Galaxies in morphological studies will be less massive or fainter than galaxies in pair studies if a uniform stellar-mass or brightness limit is adopted for sample selection. We should keep this in mind when merger rates of galaxies with disturbed morphology are directly compared with merger rates of pairs.

\subsection{Merger Rate of Tidal-Tailed Galaxies}

Using the $A_{\rm O}-D_{\rm O}$ technique together with visual examination, we identify 461 galaxies having long tidal tails among 35\,076 galaxies with  $\log (M_\star/M_\odot)\ge 9.5$ and $0.2\le z\le1$ in the COSMOS field. The tidal tails are known as low-surface-brightness features and hardly detectable at high $z$ due to the cosmic dimming effect. By redshifting a sample of model galaxies constructed from low-$z$ ones with a diversity of morphologies, we quantify the systematic biases in the measurements of the morphological parameters $A_{\rm O}$ and $D_{\rm O}$ at three redshifts $z\sim0.3, 0.5$ and 0.9. The corrections for the incompleteness in detection of tidal tails are derived through simulations and applied to the measured numbers of mergers listed in Table~\ref{tab:tab1} to obtain the merger fractions  in different stellar mass and redshift bins.  Adopting an average detection timescale of 0.5\,Gyr, the merger fractions can be converted into merger rates.  As shown in Figure~\ref{fig:figJ}, our results reveal that the merger fraction of tidal-tailed galaxies mildly increases with redshift following $\propto (1+z)^{2.0\pm0.4}$, and less massive galaxies contain a higher fraction of tailed mergers than massive ones at all cosmic epochs examined.

The increase of galaxy fraction with redshift is generally consistent with the most measurements listed in Section~\ref{sec:sum}. Moreover,  the dependence of merger fraction on stellar mass (or luminosity) are also reported in previous works (e.g., \citealt{2004ApJ...617L...9L,2005ApJ...625..621B,2009ApJ...697.1369B,2009A&A...498..379D}; B10). However, the measurements based on close pairs \citep[e.g.,][]{2009ApJ...697.1369B} favor the opposite that massive galaxies are more likely to host companions than less massive galaxies. A possible reason for this discrepancy is the bias in pair identification due to the strong clustering effect \citep[i.e., galaxies reside in groups or clusters but are assumed to merge,][]{2008MNRAS.391.1489K}. Another reason could be due to the cosmic variance as the sample used in \citet{2009ApJ...697.1369B} is rather small and probably leave large uncertainties in their results.

The results shown in Figure~\ref{fig:figJ} implies the merger fraction of massive galaxies increases more rapidly with redshift than less massive galaxies. The best-fit power index $m$ is 1.6, 2.1, 3.5 and 5.0 for galaxy mass bins from the low- to high-mass end, respectively. This predicts that merger fraction of massive galaxies would become higher than that of less massive galaxies at $z>1$, consistent with the findings from previous studies (e.g., \citealt{2008MNRAS.386..909C}; B10). Moreover, we stress that merger fraction of galaxies with $10^{9.5}\,M_{\odot}\le M_\star<10^{10}\,M_{\odot}$ at $0.8\leq z\leq 1$ may probably be underestimated due to the difficulties in finding faint tidal tails in visual examination of low-mass galaxies at $z\sim 0.9$. The underestimation of merger fraction may result in a lower $m$ for the lowest-mass bin. The sample completeness of the lowest-mass bin is relatively larger although the global completeness galaxies at $M_\star \ge 10^{9.5}\,M_{\odot}$ and $z=1$ is estimated to be about 95\%. This also increases the uncertainties of merger fraction for low-mass galaxies.

B10 used similar probes (tidal tails and bridges) to identify galaxy mergers.  We notice that their merger fraction is substantially higher and evolve more rapidly with redshift than our results. The discrepancy may be attributed to several reasons. First, more than one third of the mergers identified in B10 are not included in our statistics. These mergers either display a tidal bridge linking one galaxy to the other or double cores but no tail is presented. Second, a close pair of galaxies with tidal tails is counted as one merging event in our analysis. Such system is seen as two mergers in B10 if a bridge is shown, leading to roughly another 3\% enhancement of their merger rate. Moreover, the parent sample in B10 is selected using both stellar mass and brightness cuts. We argue that the brightness cut may lead to strong bias in particularly at high $z$. Star formation triggered by wet mergers can enhance the brightness of the host galaxies. A brightness selection would favor the detection of star-forming galaxies and possibly more mergers.


Galaxy pairs can be divided into early-type galaxy pairs, late-type galaxy pairs and mixed pairs. Mergers between two gas-poor early-type galaxies (dry mergers) are thought to be a major channel for the assembly of massive ($M_\star\ge10^{11}\,M_{\odot}$) galaxies at $z<1$ \citep{2006ApJ...640..241B,2005AJ....130.2647V,2013MNRAS.428.1460B,2014ARA&A..52..291C}. But early-type galaxy pairs contribute little to the total number of galaxy pairs, from $\sim25\%$ in the present day to $\sim 10\%$ at $z\sim1$ \citep[e.g.,][]{2009A&A...498..379D}. Figure~\ref{fig:figK} presents the comparison of the merger rate of tidal-tailed galaxies with the global merger rate derived from close galaxy pairs in the literature (see Section~\ref{sec:sum}). We find that the merger rate of tailed galaxies roughly accounts for a half of the global merger rate when the contribution of early-type galaxy pairs are removed. We conclude from Figure~\ref{fig:figK} that roughly half of disk-involved major mergers have undergone a phase exhibiting long tidal tails at $z<1$.

The absence of tidal tail in disk-involved mergers may be attributed to special orbital parameters. For instance, long tidal tails are preferentially formed in prograde encounters where the spin of a disk aligns with the orbital spin. However, only weak, warped tails can be generated in retrograde encounters. Therefore, our merger sample is biased against retrograde mergers. In the case of head-on collision (with impact parameter of zero) between a disk and a compact spheroid galaxy, ring-like debris instead of tidal tail forms and radially expands due to the density wave created by the interaction \citep{2012MNRAS.420.1158M,2012MNRAS.425.2255F}. The head-on collisions are not commonly seen and Arp\,147 \citep{2011MNRAS.417..835F} and 2MASX\,J06470249+4554022, dubbed as ``Auriga's Wheel'' \citep{2011ApJ...741...80C} are two of this kind. On the other hand, in our work tidal tails are identified if they are longer than the diameters of their hosts. This implies that tidal tails with fixed lengths are easily selected in small galaxies and thus our selection probably misses some tidal tails in larger galaxies. In addition, a tidal tail could not be identified if its orientation is nearly coincident with the line of sight and appears to be much shortened due to the projection effect.



   Tidal tails of mergers at $z>2$ might be quite different from their local counterparts. Representative disk galaxies at $z>2$ contain a few kpc-scaled blue spheroids on their disks and have clumpy morphology \citep{2004ApJ...604L..21E,2007ApJ...658..763E,2009ApJ...692...12E}. The formation of the clumpy disks might result from the gravitational instability and fragmentation in gas-rich disks which are continuously fed by cosmological streams \citep[e.g.,][]{2004A&A...413..547I,2004ApJ...611...20I,2009ApJ...703..785D}. Simulations predicted that ongoing mergers of these galaxies will form clumps and filaments in the intergalactic space instead of long tidal tails as what observed in lower redshifts \citep{2011ApJ...730....4B}. It is worth noting that the formation of tidal tails might not be common during the disk-disk mergers at $z>2$ and care should be taken when counting mergers with tidal tails in those cosmic epochs.

\section{Conclusion}

          We carry out a careful morphological analysis using deep \textit{HST}/ACS F814W imaging data for a complete sample of 35\,076 galaxies  with $M_\star \ge 10^{9.5}\,M_{\odot}$ at $0.2\leq z\leq 1$ selected from the UltraVISTA catalog of the COSMOS field.  Two newly developed morphological parameters, the outer asymmetry $A_{\rm O}$ and the outer centroid deviation $D_{\rm O}$ are measured from galaxy images to probe asymmetric features in the outskirts of galaxies.  The sample galaxies of different morphological types form a tight sequence in the $A_{\rm O}-D_{\rm O}$ diagram, showing that the degree of morphological disturbance globally increases with $A_{\rm O}$ and $D_{\rm O}$ and these with highly-irregular structures  occupy the high-value end.  Merging galaxies with long tidal tails  are pre-selected using the $A_{\rm O}-D_{\rm O}$ selection technique,  and further identified through visual examination.  This type of galaxies are believed to be major mergers involving at least one disk galaxy. How the cosmic surface brightness dimming effects on the measurements of the two parameters are quantitatively estimated through simulations and the detection completeness for tidal-tailed galaxies are derived accordingly.  We summarize our results as follows:
\begin{itemize}
           \item[1.] We find that the widely-used morphological parameter, asymmetry $A$, is significantly biased  by the noise in the outskirts of a galaxy due to an incorrect treatment of the background subtraction. We provide a revised definition of $A$ by including corrections for noise contamination. We also update the equation for computing the outer asymmetry $A_{\rm O}$ and correct for the the noise term. Details are given in the Appendix~A.
           \item[2.]  Simulations are performed to quantify the effects of the cosmic surface brightness dimming on the measurements of $A_{\rm O}$ and $D_{\rm O}$ as a function of redshift, giving that the two parameters would be overestimated for galaxies with regular morphologies (intrinsically small in $A_{\rm O}$ and $D_{\rm O}$) and be underestimated for those with disturbed morphologies (intrinsically large  in $A_{\rm O}$ and $D_{\rm O}$). The degree of the estimate bias is inversely correlated with galaxy stellar mass (and luminosity). The simulations suggest that 78, 72 and 62 percents of long tidal tails are detectable with the COSMOS \textit{HST}/ACS images at $z\sim0.5, 0.7$ and 0.9, respectively.
	   \item[3.]  In total 461 galaxies are identified to have long tidal tails from 13\,227 merger candidates selected using the $A_{\rm O}-D_{\rm O}$ technique from the parent sample of 35\,076 galaxies over $0.2\le z\le1$.   These galaxies spread in the same region as star-forming galaxies in the $U-V$ and $V-J$ diagram, confirming that tidal-tailed galaxies originate from the population of star-forming galaxies with morphologies dominated by disks.
         \item[4.]  We show that the fraction of mergers with long tidal tails evolves mildly with redshift following a power-law function $f=0.64\pm0.13\%\,(1+z)^{2.0\pm0.4}$ out to $z=1$. We also find that the tidal-tailed merger fraction is higher for less massive galaxies. However, merger fraction of massive sample evolves more strongly with redshift, indicating higher merger fraction of massive sample at $z\ge1$. These are consistent with the results from previous studies.
          \item[5.] Adopting an average detection timescale of  0.5\,Gyr for tidal-tail-traced mergers, we obtain the evolution of the merger rate of such type of merging galaxies as $0.01\pm0.007(1+z)^{2.3\pm1.4}$\,Gyr$^{-1}$.  Combined with the global merger rate derived from close pairs, we find that nearly half of disk-involved major mergers at $z<1$ have undergone a phase with long tidal tails. We infer that a present-day galaxy would undergo about 0.3 major mergers creating long tidal tails since $z=1$, roughly one-third of the total budget of major mergers.

    
    \end{itemize}



    We thank the referee for his/her constructive suggestions and comments, which greatly improve our manuscript. We thank Jeniefer Lotz  for helpful suggestions during the conference \textit{The Many Pathways to Galaxy Growth} in 2015. We also thank Cong Ma for the meaningful discussion on improving the non-parametric method.
    This work is supported by National Basic Research Program of China (973 Program 2013CB834900), the Strategic Priority Research Program ``The Emergence of Cosmological Structures'' of the Chinese Academy of Sciences (grant No. XDB09000000) and NSFC grant (U1331110).

   \newpage

   \appendix

   One challenge for the non-parametric methods is how to get rid of noise contamination. The Gini coefficient is strongly dependent on the noise level of imaging data and suffers from large uncertainties when signal-to-noise ratio is not sufficiently high \citep{2008ApJS..179..319L}. For asymmetry $A$, noise contribution is estimated by measuring $A$ on a nearby blank sky region with the same aperture. However, the noise correction for $A$ overestimates the real noise contribution \citep[see Figure 1 in][]{2009ApJ...697.1764S}. Here we present a detailed analysis on the noise effects in quantifying $A$ and introduce a revised correction for noise contamination.

\begin{figure}[] 
\centering
\includegraphics[scale=0.7]{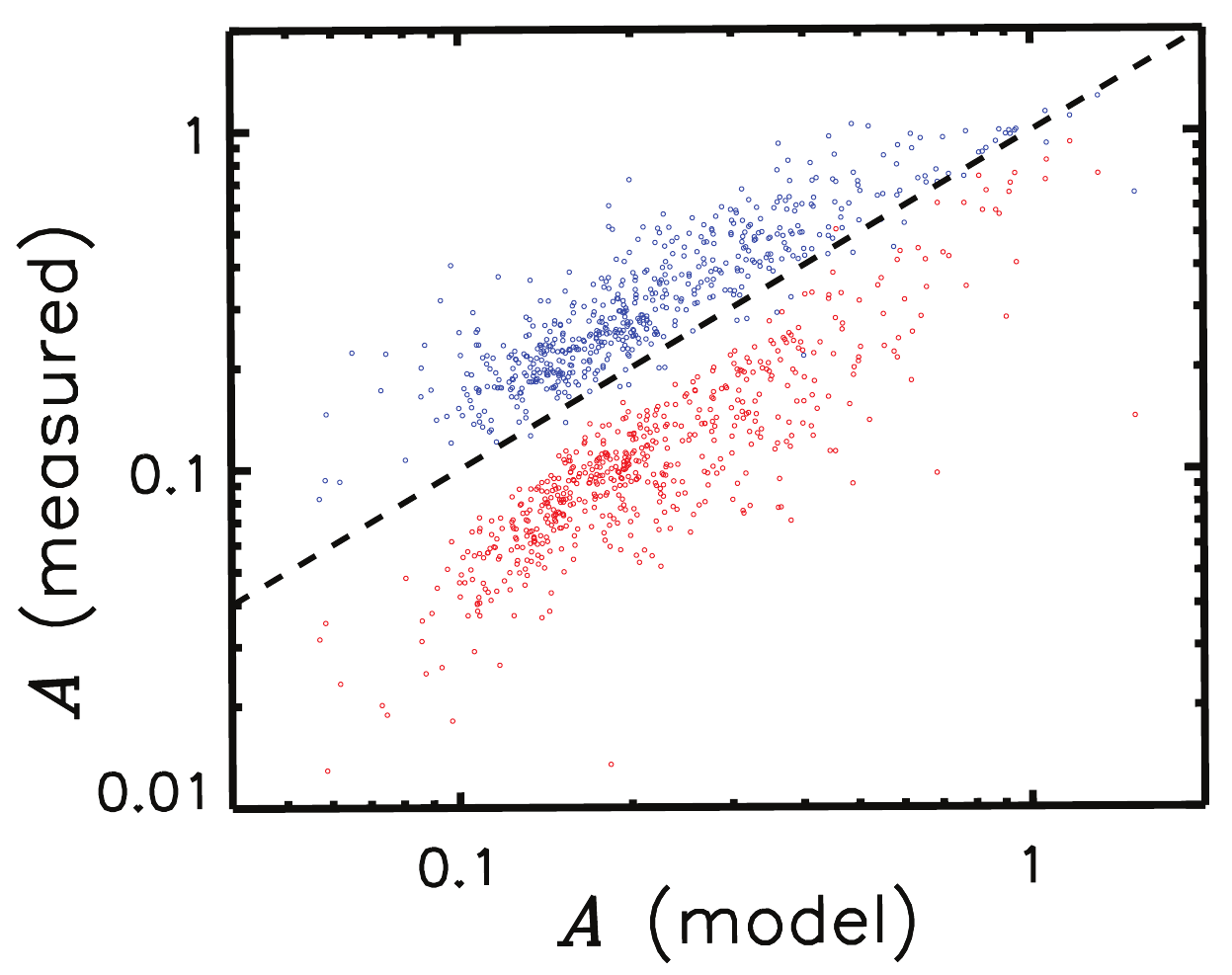}
\caption{Comparison of  asymmetry measurements from galaxy images with and without noise. Asymmetry of the noise-free model image is marked as $A(\rm model)$. For galaxy image with noise, the measurement of $A$ is biased, as denoted by the blue data points (no correction for noise contamination).  Red data points show the measurements of $A$ after including correction for the noise contamination introduced by \citet{2003ApJS..147....1C}. It is clearly that the noise correction leaves $A$ underestimated. The dashed line indicates the 1:1 relation between $A(\rm measured)$ and $A(\rm model)$.}
\label{fig:fig2}
\end{figure}

We construct a sample of model galaxies for testing the morphological parameter $A$. The sample consists of 569 isolated galaxies from \citet{2014ApJ...787..130W} (see Section~\ref{sec:simulation} for more details about model generation). An artificial image is obtained through combining a model galaxy image with a background noise image. The parameter $A$ is measured from both the artificial image and the noise-free model image following the definition given in \citet{2003ApJS..147....1C} as 
\begin{equation}\label{equ:eqOA}
         A=\frac{{\rm min}\left(\sum{|I_{0}-I_{180}|}\right)}{ \sum{|I_0|}}-\frac{{\rm min}\left(\sum{|B_{0}-B_{180}|}\right)}{ \sum{|I_0|}},
\end{equation}
where $I_{0}$ is the intensity of the galaxy and $I_{180}$ is the 180$\degr$-rotated intensity. The second term refers to the correction for noise contamination. A group of pixels are randomly selected from the background regions to form the background noise image. In the measurement, an aperture of 1.5 times the circular Petrosian aperture is adopted to estimate the noise contamination. Figure~\ref{fig:fig2} presents comparison of the measure of $A$ from noise-added and noise-free images for the 569 model galaxies. We find that noise contamination leads $A$ to be systematically overestimated and the significant scatter indicates the uncertainties of noise contribution which is about 20\% for a high-S/N galaxy and reaches 50\% for a low-S/N galaxy \citep{2009ApJ...697.1764S}. The noise correction introduced in \citet{2003ApJS..147....1C} overestimates the noise contamination and biases $A$ to be smaller. This is because that all background pixels within the Petrosian aperture are used to calculate the noise contamination. As discussed in next section, however, we argue that noise contamination is correlated with the number fraction of noise dominated pixels of a galaxy indicating that only a part of the noise contribution estimated by Equation \ref{equ:eqOA} is real. Furthermore, the measured $A$ from noise-added images is found to be lower than that from noise-free model images in a few cases even if no noise correction is adopted. This is because that noise dramatically changes the segmentation maps of the model galaxies. In our measurements, only pixels within 1.5 times the circular Petrosian aperture are counted to calculate $A$ and the extended structures of the model galaxies out of the aperture are ignored, leading to an underestimation in $A$. For a noise-free model galaxy image, all pixels associated with the galaxy are used to compute $A$.

\subsection{A Revised Noise Correction for Rotational Asymmetry}\label{sec:sec3.2}

      To calculate $A$ of a galaxy, one needs to measure the signal remained in the residual image relative to that in the scientific image. Mathematically, the modules of pixels in each image are derived and then the summation is performed. Pixels dominated by random noise likely significantly contribute to the summation even if little signal is included. As a consequence, different outputs are generated from noise-added image of a galaxy and the noise-free model. For a luminous galaxy that all pixels are brighter than $3\,\sigma$ of the noise, equivalent outputs are generated from the noise-added image and the model galaxy. It is thus important to quantify the number fraction of noise dominated pixels in the image of a galaxy against those signal dominated. In practice,1\,$\sigma$ of noise level is adopted as the criterion for the classification. The noise level is measured independently from the background.

      The updated equation of $A$ is given by
         \begin{equation} \label{equ:eq2}
         A=\frac{{\rm min}\left(\sum{|I_{0}-I_{180}|}\right)-\delta_{2}}{ \sum{|I_{0}|}-\delta_{1}},
         \end{equation} 
         where
         $\delta_{1}=f_1*\sum{|B_{0}|}$,
         $f_1=N_{flux<1\,\sigma}/N_{\rm all}$, 
         $\delta_{2}=f_2*{\rm min}\left(\sum{|B_{0}-B_{180}|}\right)$ and 
         $f_2=N_{|flux|<\sqrt{2}\,\sigma}/N_{\rm all}'$.  Parameters $\delta_{1}$ and $\delta_{2}$ measure the noise contributions within the $I_{0}$ and the residual image $I_{0}-I_{180}$ respectively. Parameters $f_{1}$ and $f_{2}$ give the number fractions of noise dominated pixels relative to the total while $N_{\rm all}$ and $N_{\rm all}'$ mark the total number of pixels assigned to the galaxy in $I_{0}$ and the residual image. Parameter $\sigma$ is the standard deviation of noise in $I_{0}$. Moreover, the residual image is derived by combining $I_{0}$ with $-I_{180}$ where image of gaussian distributed noise as well as its 180$\degr$-rotated version are overlapped on. The standard deviation of the noise in the residual image increases by a factor of $\sqrt2$ relative to that in the $I_{0}$. For this reason, pixels within $\pm\sqrt2\,\sigma$ are defined as noise dominated pixels in the residual image. 
 
     The size of a galaxy is determined by the Petrosian aperture and the pixels within 1.5 times the elliptical Petrosian aperture of the galaxy are used to compute $A$. Elliptical apertures more closely follow the light profile of galaxies than circular apertures, especially for edge-on disks. The rotational centers of either model galaxies, background images or artificial galaxies are determined by an iterative process respectively. The initial center (usually the flux-weighted centroid of a galaxy) together with the eight pixels surrounded will be tested in turn until the minimum value of asymmetry is found.

\begin{figure*}[] 
\centering
\includegraphics[width=0.5\textwidth]{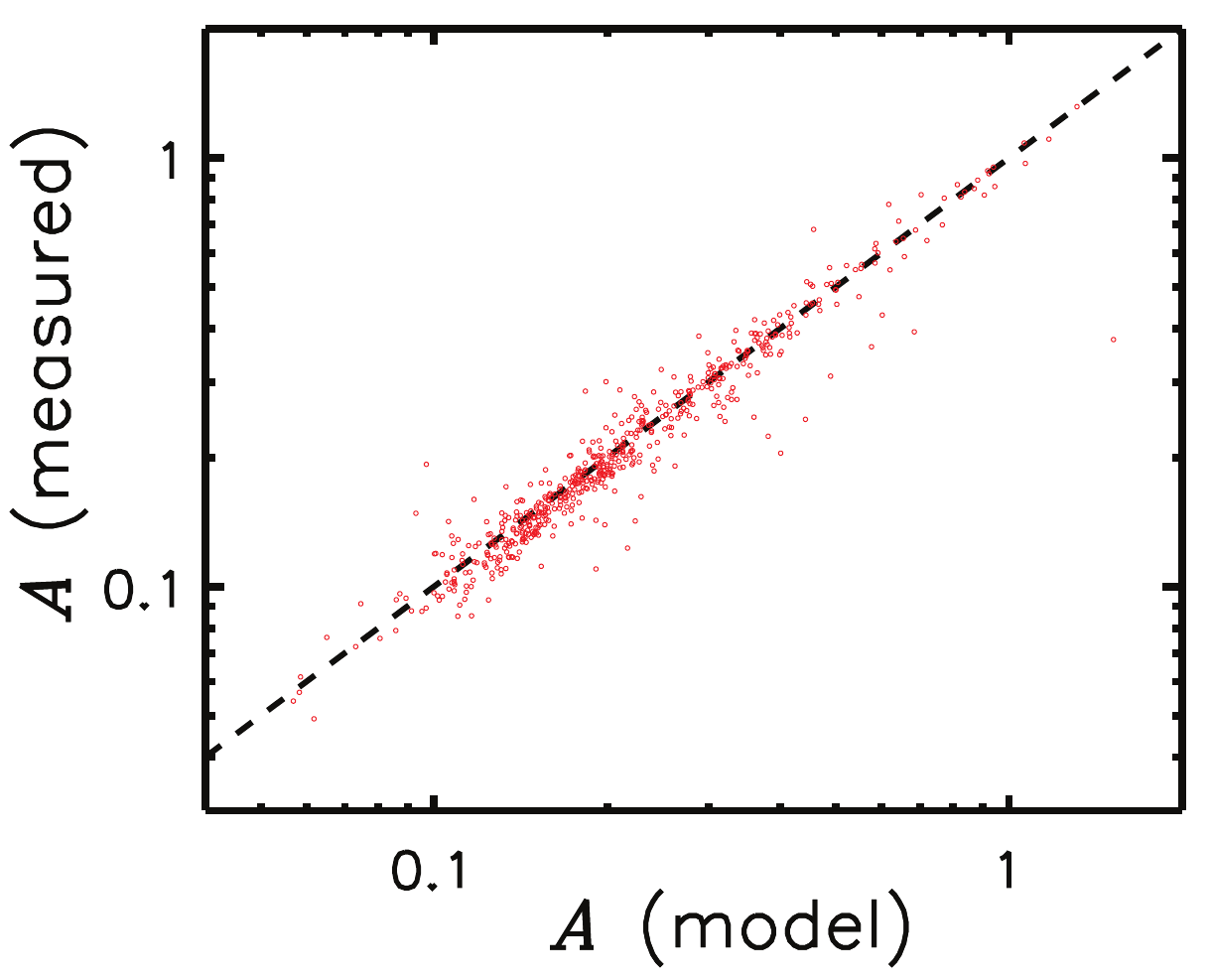}
\caption{This figure compares the new measurements of $A$ from noise-added images with the measurements from the models. The dashed line indicates the 1:1 relation between $A(\rm measured)$ and $A(\rm model)$.}
\label{fig:fig3}
\end{figure*}

       Figure~\ref{fig:fig3} compares the new measurements of $A$ from noise-added images with the measurements from the models. The new measurements of $A$ are in line with the initial values of the models. Only $<5\%$ cases show large offsets from the initial $A$. In these cases, extended structures out of the 1.5 times elliptical Petrosian radius of an artificial galaxy are not included in the measurements but they are included when the initial $A$ of the model is measured . The determination of rotational center of the artificial galaxy thus deviates from that of the model galaxy. We also notice that the slope of the data-points in Figure~\ref{fig:fig3} will change if the criterion for selecting noise dominated pixels in the $I_{0}$ is set other than $1\,\sigma$ of the noise.

      We thus apply the revised noise correction to $A_{\rm O}$. The details for updated definition of $A_{\rm O}$ is given in Section~\ref{sec:Ao}. In contrast to the Petrosian radius, the size of a galaxy is directly determined by the segmentation map when $A_{\rm O}$ is computed. We find that our new method for noise correction is reliable when $A$ or $A_{\rm O}$ of a galaxy is measured  and thus is independent on the size of the galaxy.





\end{CJK*}
\end{document}